\documentclass[english, pre, twocolumn, showpacs]{revtex4}
\usepackage[T1]{fontenc}
\usepackage[latin9]{inputenc}
\usepackage{verbatim}
\usepackage{amsmath}
\usepackage{amssymb}
\usepackage{graphicx}
\usepackage{babel}

\begin{document}

\preprint{This line only printed with preprint option}

\title{Numerical formulation of three-dimensional scattering problems for optical structures}

\author{Tatsuya Usuki}

\email{t-usuki@petra-jp.org}

\homepage{http://www.petra-jp.org/}

\selectlanguage{english}

\affiliation{Institute for Photonics-Electronics Convergence System Technology (PECST), Photonics Electronics Technology Research Association (PETRA), West 7 SCR, 16-1 Onogawa, Tsukuba, Ibaraki 305-8569}

\date{\today}
\begin{abstract}
This paper describes a numerical formulation for calculating wave propagation with high precision in a three-dimensional system.
Yee's discretization scheme is used to formulate a frequency domain method that is compatible with the finite-difference time-domain (FDTD) procedure.
When the S-matrix satisfies a unitarity (power flow conservation) condition, the method enables arbitrary S-matrix elements to be obtained within a numerical error of less than $10^{-8}$ ($2 \times 10^{-13}$) for double precision format.
\end{abstract}
\pacs{02.60.Cb,41.20.Jb,42.15.Eq,42.25.Bs}
\maketitle

\section{introduction}
Numerical simulations are important for studying electromagnetic-wave propagation in optical physics~\cite{Joannopoulos} and for designing silicon photonics~\cite{Kimerling, Miller}.
One of the most successful numerical methods is the finite-difference time-domain (FDTD) method~\cite{198008_taflove}.
It is suitable for visualizing dynamical propagation of electromagnetic waves.
Simulations should not only give us a general understanding of the propagation, but also the details of the optical scattering. 
\\ \indent
Designing chips such as silicon optical interposers~\cite{Urino} requires highly precise simulations including the transmittance and reflectance of the fundamental and higher order modes at each wavelength (i.e., each frequency).
Small reflections may cause substantial instability~\cite{Yamamoto} between devices on the optical chip, and the small losses that result may build up (e.g. see Table I in Ref.~\cite{Urino}).
Thus, we need a way to confirm that the numerical results are precise when the scattering properties are simulated. 
\\ \indent
Here, we should note that the precision of a numerical calculation, which is affected by both numerical method (e.g. numerical stability for the FDTD~\cite{Taflove_book}) and numerical implementation (e.g. floating-point arithmetic~\cite{IEEE754}), is essentially different from the accuracy of numerical modeling that includes both the choice of the fundamental equation (e.g. microscopic nonlocal approach~\cite{Cho} is one such choice) and the discretization of the  numerical procedure (e.g. numerical dispersion for the FDTD~\cite{Taflove_book}).
In a scattering simulation, the error related to the accuracy is often explicit and predictable, but the error related to the precision is apt to be implicit and unforeseeable.
The S-matrix approach~\cite{Wheeler} is widely used to study scattering problems~\cite{Buttiker}, and numerical S-matrices have already been applied to scattering simulations on photonic crystal slabs~\cite{Tikhodeev,Li,Liscidini} and metal films~\cite{Anttu}.
Unfortunately, no method as yet has been discussed related to highly precise simulations in the three-dimensional optical structures.
\\ \indent
This paper proposes a numerical method that can produce precise S-matrices for designing the silicon photonics devices.
The method exploits a numerical procedure for quantum transport~\cite{Usuki, Akis}.
The numerical precision of the method is evaluated in terms of the S-matrix properties.

\section{Formulation}
\begin{figure}[ht]
\begin{centering}
\includegraphics[width=0.7\columnwidth]{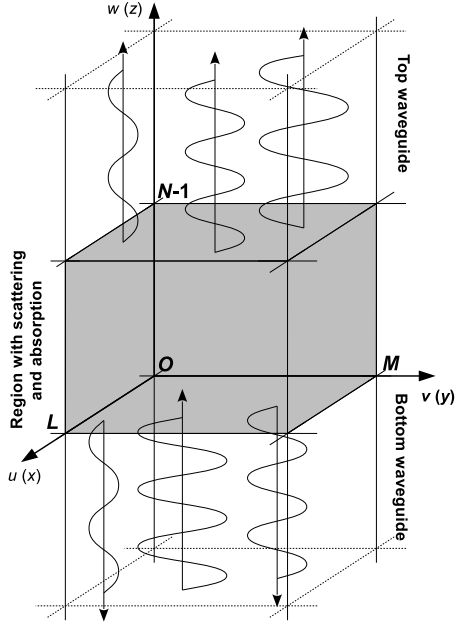}
\end{centering}
\caption{Optical system for multi-mode scattering. The propagation region  is divided into three parts.}
\label{fig: Optical-system}
\end{figure}

Consider the macroscopic Maxwell equations in the angular frequency domain ($\omega$ space),
\begin{equation}
	\begin{split}
		\nabla\times\boldsymbol{H} & = -i \omega \varepsilon_{0} \varepsilon \left(\boldsymbol{x}\right) \boldsymbol{E}\,,\\
		\nabla\times\boldsymbol{E} & =  i \omega \mu_{0} \mu \left(\boldsymbol{x}\right) \boldsymbol{H}\:,
	\end{split}
	\label{eq: Maxwell eqs omega}
\end{equation}
where $\varepsilon_{0}$ ($\mu_{0}$) is vacuum permittivity (permeability). The symbol $i$ denotes an imaginary number, and the notation ``$\exp \left( - i \omega t \right)$'' describes a harmonic oscillation.
The symbols $j$, $k$, $l$, $m$, and $n$ in the following formulation denote integers.
Instead of using dipole moments in the optical media, Eqs. (\ref{eq: Maxwell eqs omega}) are used to express the relative permittivity $\varepsilon\left(\boldsymbol{x}\right)$ and relative permeability $\mu\left(\boldsymbol{x}\right)$.
Note that $\mathrm{Im}\, \varepsilon \left(\boldsymbol{x}\right),\,\mathrm{Im} \, \mu\left(\boldsymbol{x}\right) \geq 0$
for absorbing media.
\\ \indent
The optical system in Fig. \ref{fig: Optical-system} consists of three parts: two ideal waveguides and a region with scattering and absorption.
The coordinates in Fig. \ref{fig: Optical-system} are ones transformed using $\boldsymbol{x} = \left(x,\, y,\, z\right)\rightarrow\left(u,\, v,\, w\right)$, and they are used to apply a non-uniform mesh to Eqs. (\ref{eq: Maxwell eqs omega}).
Note that the Jacobian matrix of this transformation is diagonal in order to simplify the discussion in the following subsections ${\bf A}$-${\bf F}$.

\subsection{Discrete representation}
Let us discretize the transformed space: $\left(u,\, v,\, w\right)\rightarrow\left(l,\, m,\, n\right)$, and let us use cells in Yee's lattice~\cite{Yee}. 
\begin{figure}[ht]
	\begin{centering}
		\includegraphics[width=0.6\columnwidth]{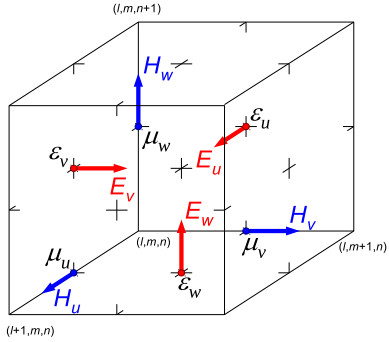}
	\end{centering}
	\caption{A $\left(l,\, m,\, n\right)$ cell in Yee's lattice.}
	\label{fig: Yee cell}
\end{figure}
 Figure \ref{fig: Yee cell} shows an arrangement of discretized
functions ($\varepsilon$, $\mu$, $\boldsymbol{E}$ and $\boldsymbol{H}$)
that are allocated a cell address $\left(l,\, m,\, n\right)$.
For example, the $u$ component of the magnetic field $H_{u}\left(l,\, m,\, n\right)$
means $H_{u}$ at $u=l+1/2$, $v=m$, $w=n$ in the figure.
The electromagnetic field in $\left(x,\, y,\, z\right)$ coordinates is related to that of the $\left(u,\, v,\, w\right)$ coordinates in the following manner.
\begin{equation}
	\begin{split}\sqrt{\mu_{0}}H_{x} & = f_{1}\left(l\right)\, g_{0}\left(m\right)\, h_{0}\left(n\right)\, H_{u}\left(l,\, m,\, n\right),\\
		\sqrt{\mu_{0}}H_{y} & =f_{0}\left(l\right)\, g_{1}\left(m\right)\, h_{0}\left(n\right)\, H_{v}\left(l,\, m,\, n\right),\\
		\sqrt{\mu_{0}}H_{z} & =f_{0}\left(l\right)\, g_{0}\left(m\right)\, h_{1}\left(n\right)\, H_{w}\left(l,\, m,\, n\right),
	\end{split}
	\label{eq: Hu Hv Hw}
\end{equation}
and
\begin{equation}
	\begin{split}\sqrt{\varepsilon_{0}}E_{x} & =f_{0}\left(l\right)\, g_{1}\left(m\right)\, h_{1}\left(n\right)\, E_{u}\left(l,\, m,\, n\right),\\
		\sqrt{\varepsilon_{0}}E_{y} & =f_{1}\left(l\right)\, g_{0}\left(m\right)\, h_{1}\left(n\right)\, E_{v}\left(l,\, m,\, n\right),\\
		\sqrt{\varepsilon_{0}}E_{z} & =f_{1}\left(l\right)\, g_{1}\left(m\right)\, h_{0}\left(n\right)\, E_{w}\left(l,\, m,\, n\right),
	\end{split}
	\label{eq: Ev Ew}
\end{equation}
where $f_{k}$, $g_{k}$, and $h_{k}$ are defined as 
\begin{equation}
	\begin{split}
		f_{k}^{-2}\left(l\right) & =\left. \frac{\omega}{c} \frac{dx}{du}\right|_{u=l+k/2}\,,\\
		g_{k}^{-2}\left(m\right) & =\left. \frac{\omega}{c} \frac{dy}{dv}\right|_{v=m+k/2}\,,\\
		h_{k}^{-2}\left(n\right) & =\left. \frac{\omega}{c} \frac{dz}{dw}\right|_{w=n+k/2}\,.
	\end{split}
	\label{eq: fj gj hj}
\end{equation}
Here, $c = 1/\sqrt{\varepsilon_{0}\mu_{0}}$ is the velocity of light in a vacuum. We make the domain of variables $u$ and $v$ finite: $0\leq u<L$ and $0\leq v<M$, where the boundaries $L$ and $M$ are integers. 
Furthermore, $\varepsilon_{\eta}$ and $\mu_{\eta}$ for $\eta=u,\, v,\, w$ in
Fig. \ref{fig: Yee cell} are defined as
\[
	\begin{split}\varepsilon_{u}\left(l,\, m,\, n\right) & =\left.\varepsilon_{\,}\right|_{u=l,\, v=m+1/2,\, w=n+1/2}\,,\\
	\varepsilon_{v}\left(l,\, m,\, n\right) & =\left.\varepsilon_{\,}\right|_{u=l+1/2,\, v=m,\, w=n+1/2}\,,\\
	\varepsilon_{w}\left(l,\, m,\, n\right) & =\left.\varepsilon_{\,}\right|_{u=l+1/2,\, v=m+1/2,\, w=n}\,,
	\end{split}
\]
and
\[
	\begin{split}\mu_{u}\left(l,\, m,\, n\right) & =\left.\mu_{\,}\right|_{u=l+1/2,\, v=m,\, w=n}\,,\\
		\mu_{v}\left(l,\, m,\, n\right) & =\left.\mu_{\,}\right|_{u=l,\, v=m+1/2,\, w=n}\,,\\
		\mu_{w}\left(l,\, m,\, n\right) & =\left.\mu_{\,}\right|_{u=l,\, v=m,\, w=n+1/2}\,.
	\end{split}
\]
The relative permittivity $\varepsilon_{\eta}$ and relative permeability
$\mu_{\eta}$ for $\eta=u,\, v,\, w$ satisfy periodic conditions, i.e., $\varepsilon_{\eta}\left(l+L,\, m,\, n\right)=\varepsilon_{\eta}\left(l,\, m,\, n\right)$ and $\varepsilon_{\eta}\left(l,\, m+M,\, n\right) = \varepsilon_{\eta}\left(l,\, m,\, n\right)$.
\begin{figure}[ht]
	\begin{centering}
		\includegraphics[width=0.7\columnwidth]{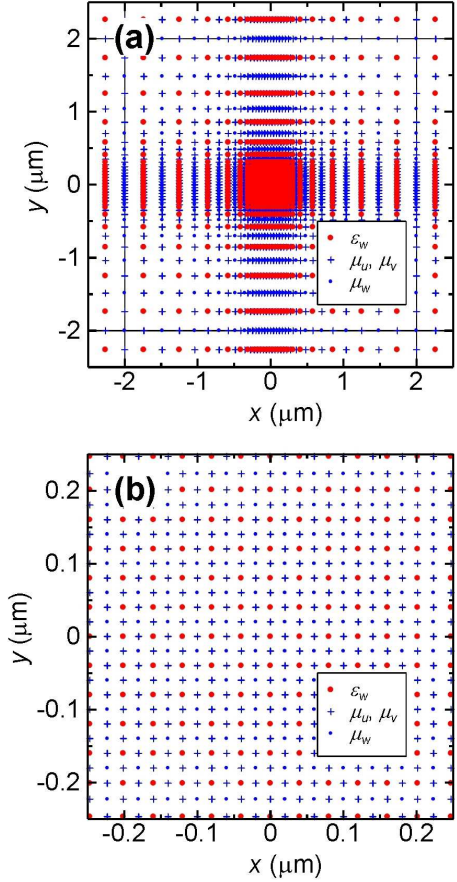}
	\end{centering}
	\caption{Grid of Yee's lattice in periodic $xy$ space:
		$x\left(L\right)-x\left(0\right)=y\left(M\right)-y\left(0\right)=4\,\mathrm{\mu m}$
		when $L=M=25$,
		$\min\left(dx/du\right)=\min\left(dy/dv\right)=40\,\mathrm{nm}$,
		 and $K_{u}=K_{v}=5$
		(see Eq. (\ref{eq: xy example}) in Appendix \ref{sec:Appendix Non_uniform_mesh}).
		(a) Grid with periodic boundary conditions (dot-broken lines).
		(b) Grid around the origin.
	}
	\label{fig: Grid}
\end{figure}
Figure \ref{fig: Grid} shows the coordinate transformation $\left(x,\, y\right)\rightarrow\left(u,\, v\right)$ described by Eqs. (\ref{eq: xy example}) in Appendix \ref{sec:Appendix Non_uniform_mesh}. 

The six components of the electromagnetic field satisfy the following conditions:
\begin{equation*}
	\begin{split}\mathcal{G}\left(l+L,\, m,\, n\right) & =B_{u}\mathcal{G}\left(l,\, m,\, n\right),\\
		\mathcal{G}\left(l,\, m+M,\, n\right) & =B_{v}\mathcal{G}\left(l,\, m,\, n\right),
	\end{split}
\end{equation*}
where $\mathcal{G}=H_{\eta}$, $E_{\eta}$ for $\eta=u,\, v,\, w$.
The parameters $B_{u}$, $B_{v}$ are generally complex numbers, and they satisfy $ \left| B_{u} \right| = \left| B_{v} \right| = 1 $.
In this paper, $B_{u} = B_{v} = 1$.
Now let us introduce the forward difference operators,
\begin{equation}
	\begin{split}\Delta_{u}\mathcal{G}\left(l,\, m,\, n\right) & \triangleq\mathcal{G}\left(l+1,\, m,\, n\right)-\mathcal{G}\left(l,\, m,\, n\right),\\
		\Delta_{v}\mathcal{G}\left(l,\, m,\, n\right) & \triangleq\mathcal{G}\left(l,\, m+1,\, n\right)-\mathcal{G}\left(l,\, m,\, n\right),\\
		\Delta_{w}\mathcal{G}\left(l,\, m,\, n\right) & \triangleq\mathcal{G}\left(l,\, m,\, n+1\right)-\mathcal{G}\left(l,\, m,\, n\right).
	\end{split}
	\label{eq: difference operator}
\end{equation}
At $u = L-1$ and $v = M-1$, $\Delta_{u}$ and $\Delta_{v}$ are defined as
\[
	\begin{split}\Delta_{u}\mathcal{G}\left(L-1,\, m,\, n\right) & \triangleq B_{u}\mathcal{G}\left(0,\, m,\, n\right)-\mathcal{G}\left(L-1,\, m,\, n\right),\\
		\Delta_{v}\mathcal{G}\left(l,\, M-1,\, n\right) & \triangleq B_{v}\mathcal{G}\left(l,\,0,\, n\right)-\mathcal{G}\left(l,\, M-1,\, n\right).
	\end{split}
\]
The backward difference operator is $-\Delta_{u}^{\mathrm{T}}$, where ``$^{\mathrm{T}}$'' denotes the transpose.
Using Eqs. (\ref{eq: fj gj hj}) and (\ref{eq: difference operator}), we can define modified difference operators:
\begin{equation}
	\tilde{\Delta}_{u}=f_{1}\Delta_{u}f_{0},\;\tilde{\Delta}_{v}=g_{1}\Delta_{v}g_{0},\;\tilde{\Delta}_{w}=h_{1}\Delta_{w}h_{0}.\label{eq: modified tilde diff}
\end{equation}
From Eqs. (\ref{eq: Maxwell eqs omega}) and (\ref{eq: modified tilde diff}),
the $u$ and $v$ components of the electromagnetic field satisfy
\begin{equation}
	\begin{split}\tilde{\Delta}_{w}\boldsymbol{H}_{uv}\left(n\right) & =i\boldsymbol{M}_{HE}\left(n\right)\boldsymbol{E}_{vu}\left(n\right),\\
		-\tilde{\Delta}_{w}^{\mathrm{T}}\boldsymbol{E}_{vu}\left(n\right) & =i\boldsymbol{M}_{EH}\left(n\right)\boldsymbol{H}_{uv}\left(n\right),
	\end{split}
	\label{eq: mod Maxwell HE and EH}
\end{equation}
with the $2LM\times2LM$ matrices,
\begin{equation}
	\begin{split}
		\boldsymbol{M}_{HE} & =\left(
		\begin{array}{cc}
			- \tilde{\Delta}_{u} \boldsymbol{\mu}_{w}^{-1} \tilde{\Delta}_{u}^{\mathrm{T}} + \boldsymbol{\varepsilon}_{v}
			& -\tilde{\Delta}_{u} \boldsymbol{\mu}_{w}^{-1} \tilde{\Delta}_{v}^{\mathrm{T}}\\
			- \tilde{\Delta}_{v} \boldsymbol{\mu}_{w}^{-1} \tilde{\Delta}_{u}^{\mathrm{T}}
			& - \tilde{\Delta}_{v} \boldsymbol{\mu}_{w}^{-1} \tilde{\Delta}_{v}^{\mathrm{T}} + \boldsymbol{\varepsilon}_{u}
		\end{array}
		\right),\\
		\boldsymbol{M}_{EH} & =\left(
		\begin{array}{cc}
			- \tilde{\Delta}_{v}^{\mathrm{T}} \boldsymbol{\varepsilon}_{w}^{-1} \tilde{\Delta}_{v} + \boldsymbol{\mu}_{u}
			& \tilde{\Delta}_{v}^{\mathrm{T}} \boldsymbol{\varepsilon}_{w}^{-1} \tilde{\Delta}_{u}\\
			\tilde{\Delta}_{u}^{\mathrm{T}} \boldsymbol{\varepsilon}_{w}^{-1} \tilde{\Delta}_{v}
			& - \tilde{\Delta}_{u}^{\mathrm{T}} \boldsymbol{\varepsilon}_{w}^{-1} \tilde{\Delta}_{u} + \boldsymbol{\mu}_{v}
		\end{array}
		\right),
\end{split}
\label{eq: mod MHE MEH}
\end{equation}
and the $LM \times LM$ diagonal matrices, 
\[
	\begin{split}
		\boldsymbol{\varepsilon}_{\eta} & = \mathrm{diag} \left(
		\begin{array}{ccc}
			\varepsilon_{\eta}\left(0,\,0,\, n\right),
			& \ldots\,, & \varepsilon_{\eta}\left(L-1,\, M-1,\, n\right)
		\end{array}
		\right),\\
		\boldsymbol{\mu}_{\eta} & = \mathrm{diag} \left(
		\begin{array}{ccc}
			\mu_{\eta}\left(0,\,0,\, n\right), & \ldots\,, & \mu_{\eta}\left(L-1,\, M-1,\, n\right)
		\end{array}
		\right).
\end{split}
\]
The $2LM\times1$ column vectors $\boldsymbol{H}_{uv}$ and $\boldsymbol{E}_{vu}$ in Eq. (\ref{eq: mod Maxwell HE and EH}) are expressed  as
\[
	\boldsymbol{H}_{uv} = \left(
	\begin{array}{c}
		\boldsymbol{H}_{u}\\
		\boldsymbol{H}_{v}
	\end{array}
	\right),\quad \boldsymbol{E}_{vu} = \left(
	\begin{array}{c}
		- \boldsymbol{E}_{v}\\
		\boldsymbol{E}_{u}
	\end{array}
	\right).
\]
Here,
\[
	\begin{split}
		\boldsymbol{H}_{\eta} & = \left(H_{\eta}\left(0,\,0,\, n\right)\,\cdots\, H_{\eta}\left(L-1,\, M-1,\, n\right)\right)^{\mathrm{T}},\\
		\boldsymbol{E}_{\eta} & =\left(E_{\eta}\left(0,\,0,\, n\right)\,\cdots\, E_{\eta}\left(L-1,\, M-1,\, n\right)\right)^{\mathrm{T}}.
	\end{split}
\]
From Eqs. (\ref{eq: Maxwell eqs omega}), the $w$-components $H_{w}$
in Eqs.(\ref{eq: Hu Hv Hw}) and $E_{w}$ in Eqs.(\ref{eq: Ev Ew}) can be expressed as 
\[
	\begin{split}
		\boldsymbol{E}_{w}\left(n\right) & = \frac{i}{\boldsymbol{\boldsymbol{\varepsilon}}_{w}\left(n\right)}
		\left(\tilde{\Delta}_{u}\boldsymbol{H}_{v}\left(n\right)-\tilde{\Delta}_{v}\boldsymbol{H}_{u}\left(n\right)\right),\\
		\boldsymbol{H}_{w}\left(n\right) & = \frac{i}{\boldsymbol{\boldsymbol{\mu}}_{w}\left(n\right)}
		\left(\tilde{\Delta}_{u}^{\mathrm{T}}\boldsymbol{E}_{v}\left(n\right)-\tilde{\Delta}_{v}^{\mathrm{T}}\boldsymbol{E}_{u}\left(n\right)\right)\,.
	\end{split}
\]

\subsection{Wave propagation in ideal waveguides}
\begin{figure}[ht]
	\begin{centering}
		\includegraphics[width=0.9\linewidth]{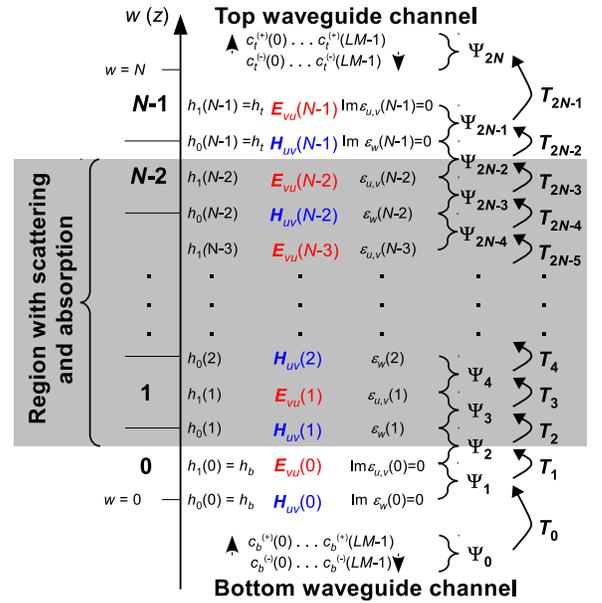}
	\end{centering}
	\caption{Graphic depiction of Eqs. (\ref{eq: imaginary 0})-(\ref{eq: T 2N-1 half}).}
	\label{fig: Transfer-matrix.}
\end{figure}

For the bottom ideal waveguide, $\boldsymbol{M}_{HE}$, $\boldsymbol{M}_{EH}$ of Eqs. (\ref{eq: mod MHE MEH}) and $h_{j}$ in Eqs. (\ref{eq: fj gj hj}) are
\[
	\begin{split}
		\boldsymbol{M}_{HE}\left(n\right) & = \boldsymbol{M}_{bHE},\;\boldsymbol{M}_{EH}\left(n\right) = \boldsymbol{M}_{bEH},\\
		h_{0}\left(n\right) & = h_{1}\left(n\right) = h_{b},\;\mathrm{as}\; n\leq0.
	\end{split}
\]
For the top ideal waveguide, they are
\[
	\begin{split}
		\boldsymbol{M}_{HE}\left(n\right) & = \boldsymbol{M}_{tHE},\;\boldsymbol{M}_{EH}\left(n\right)=\boldsymbol{M}_{tEH},\\
		h_{0}\left(n\right) & = h_{1}\left(n\right) = h_{t},\;\mathrm{as}\; n\geq N-1.
	\end{split}
\]
Since the permittivity in the ideal waveguides is real, we have
\begin{equation}
	\begin{split}
		\mathrm{Im}\boldsymbol{M}_{\kappa HE} & =\mathrm{Im}\boldsymbol{M}_{\kappa EH}=0
	\end{split}
	\label{eq: imaginary 0}
\end{equation}
for $\kappa=b,\, t$. Figure \ref{fig: Transfer-matrix.} depicts the arrangement of the above equations.
The eigenvalue equation for the optical modes of the ideal waveguides is
\begin{equation}
	\begin{split}
		\boldsymbol{M}_{\kappa HE}\boldsymbol{M}_{\kappa EH}\boldsymbol{u}_{\kappa}\left(j\right)
		& =\Lambda_{\kappa}^{2}\left(j\right)\boldsymbol{u}_{\kappa}\left(j\right),
	\end{split}
\label{eq: eigenmode}
\end{equation}
for $0\leq j<LM$.
Here, $\boldsymbol{u}_{\kappa}$ are eigenvectors.
Appendix \ref{sec: Appendix Orthogonality_of_eigenmode} show that all eigenvectors satisfy
\begin{equation}
	\boldsymbol{u}_{\kappa}^{\mathrm{T}} \left(j\right) \boldsymbol{M}_{\kappa EH} \boldsymbol{u}_{\kappa} \left(j'\right) - \delta_{jj^{'}} = 0\,,
	\label{eq: orthogonality}
\end{equation}
where $\delta_{jj^{'}}$ is Kronecker's delta.
The propagation constants $\beta_{b}$ and $\beta_{t}$ are given by
\begin{equation}
	\begin{split}
		\beta_{\kappa}\left(j\right) & = 2\arcsin\frac{\Lambda_{\kappa}\left(j\right)}{2h_{\kappa}^{2}}\;\mathrm{for}\;\kappa=b,\, t\,,
	\end{split}
	\label{eq: propagation constant}
\end{equation}
and $0\leq\arg\beta_{\kappa}\left(j\right)<\pi$.
We use an integer $J_{\kappa}$ to separate the propagating modes and evanescent modes of the above $\beta_{\kappa}$: $\mathrm{Im}\beta_{\kappa}\left(j\right)=0$
as $0\leq j<J_{\kappa}$, and $\mathrm{Im}\beta_{\kappa}\left(j\right)\neq0$
as $J_{\kappa}\leq j<LM$.
We build a square matrix consisting of the eigenmodes $\boldsymbol{u}_{\kappa}$ of Eq. (\ref{eq: eigenmode}), a diagonal matrix $\boldsymbol{\theta}_{\kappa}$ from Eq. (\ref{eq: propagation constant}) consisting of the phases of the modes, and column vectors $\boldsymbol{\psi}_{\kappa}^{\left(\pm\right)}$ consisting of the coefficients $c_{\kappa}^{\left(\pm\right)}\left(j\right)$ of the $j$-th mode in the waveguides:
\begin{equation}
	\begin{split}
		\boldsymbol{U}_{\kappa} & = \left(\boldsymbol{u}_{\kappa}\left(0\right)\,\cdots\,\boldsymbol{u}_{\kappa}\left(LM-1\right)\right),\\
		\boldsymbol{\theta}_{\kappa}= & \mathrm{diag}\left(
		\begin{array}{ccc}
			\exp\left(i\beta_{\kappa}\left(0\right)\right), & \ldots\,, & \exp\left(i\beta_{\kappa}\left(LM-1\right)\right)
		\end{array}
		\right),\\
		\boldsymbol{\psi}_{\kappa}^{\left(\pm\right)} & = \left(c_{\kappa}^{\left(\pm\right)}\left(0\right)\,\cdots\, c_{\kappa}^{\left(\pm\right)}\left(LM-1\right)\right)^{\mathrm{T}}\,.
\end{split}
\label{eq: U theta Psi}
\end{equation}
Note that $\boldsymbol{U}_{\kappa}^{\mathrm{T}}\boldsymbol{M}_{\kappa EH}\boldsymbol{U}_{\kappa}=\boldsymbol{1}$ from Eq. (\ref{eq: orthogonality}).
The $\boldsymbol{H}_{uv}\left(0\right)$ and $\boldsymbol{E}_{vu}\left(0\right)$ in the bottom ideal waveguide are defined by using Eq. (\ref{eq: U theta Psi}):
\begin{equation}
	\begin{split}
		\boldsymbol{H}_{uv}\left(0\right)
		& = \boldsymbol{U}_{b} \left(\boldsymbol{\psi}_{b}^{\left(+\right)}+\boldsymbol{\psi}_{b}^{\left(-\right)}\right),\\
		\boldsymbol{E}_{vu}\left(0\right)
		& = \frac{i}{h_{b}^{2}}\boldsymbol{M}_{b EH}\boldsymbol{U}_{b}\\
		& \quad \times\left(\frac{1}{1-\boldsymbol{\theta}_{b}^{-1}}\boldsymbol{\psi}_{b}^{\left(+\right)}
		+\frac{1}{1-\boldsymbol{\theta}_{b}}\boldsymbol{\psi}_{b}^{\left(-\right)}\right).
	\end{split}
	\label{eq: waveguide representation}
\end{equation}
The $\boldsymbol{H}_{uv}\left(N-1\right)$ and $\boldsymbol{E}_{vu}\left(N-1\right)$ in the top ideal waveguide are defined in the same manner.

\subsection{Power flow with absorption}

Let us define the discretized formulation of time averaged power flow\cite{Okamoto}:
\begin{equation}
\begin{split}P_{z}\left(n\right) & \triangleq \frac{c^{3}}{2 \omega^{2}} \mathrm{Re}\left(\boldsymbol{E}_{vu}^{\dagger}\left(n\right)h_{1}\left(n\right)h_{0}\left(n\right)\boldsymbol{H}_{uv}\left(n\right)\right)\,,\end{split}
\label{eq: power flow}
\end{equation}
where ``$^{\dagger}$'' denotes the Hermitian conjugate.
From Eqs. (\ref{eq: propagation constant}), (\ref{eq: U theta Psi}) and (\ref{eq: waveguide representation}), the power flows $P_{bz}=P_{z}\left(0\right)$ and $P_{tz}=P_{z}\left(N-1\right)$ for Eq. (\ref{eq: power flow}) are
given by
\begin{equation}
\begin{split}P_{\kappa z} & =\sum_{j=0}^{J_{\kappa}-1}\gamma_{\kappa}^{2}\left(j\right)\left(\left|c_{\kappa}^{\left(+\right)}\left(j\right)\right|^{2}-\left|c_{\kappa}^{\left(-\right)}\left(j\right)\right|^{2}\right),\\
\gamma_{\kappa}\left(j\right) & =\sqrt{\frac{c^{3}}{4 \omega^{2}} \cot\left(\frac{\beta_{\kappa}\left(j\right)}{2}\right)}\,,
\end{split}
\label{eq: gamma}
\end{equation}
where $\gamma_{\kappa}$ is real and positive. Equations (\ref{eq: mod Maxwell HE and EH}),
(\ref{eq: power flow}), and (\ref{eq: gamma}) lead to the following relation between power flow and absorption:
\begin{eqnarray}
 &  & P_{bz}-P_{tz}\nonumber \\
 & = & \frac{c^{3}}{2 \omega^{2}} \sum_{n=1}^{N-2}\left[\boldsymbol{E}_{vu}^{\dagger}\left(n\right)\left(\mathrm{Im}\boldsymbol{M}_{HE}\left(n\right)\right)\boldsymbol{E}_{vu}\left(n\right)\right.\nonumber \\
 &  & \qquad\quad+\left.\boldsymbol{H}_{uv}^{\dagger}\left(n\right)\left(\mathrm{Im}\boldsymbol{M}_{EH}\left(n\right)\right)\boldsymbol{H}_{uv}\left(n\right)\right]\label{eq: power flow and dissipation}\\
 & = & \frac{c^{3}}{2 \omega^{2}} \sum_{n=1}^{N-2}\sum_{\eta=u,v,w}\left[\boldsymbol{E}_{\eta}^{\dagger}\left(n\right)\left(\mathrm{Im}\boldsymbol{\varepsilon}_{\eta}\left(n\right)\right)\boldsymbol{E}_{\eta}\left(n\right)\right.\nonumber \\
 &  & \qquad\qquad+\left.\boldsymbol{H}_{\eta}^{\dagger}\left(n\right)\left(\mathrm{Im}\boldsymbol{\mu}_{\eta}\left(n\right)\right)\boldsymbol{H}_{\eta}\left(n\right)\right]\,.\nonumber 
\end{eqnarray}

\subsection{Transfer matrices}

We make $4LM\times1$ column vectors $\boldsymbol{\Psi}_{k}$
of the electromagnetic modes and $4LM\times4LM$ matrices $\boldsymbol{T}_{k}$, which we call transfer matrices: $\boldsymbol{\Psi}_{k+1}=\boldsymbol{T}_{k}\boldsymbol{\Psi}_{k}$. The $\boldsymbol{\Psi}_{k}$ are defined as 
\[
\begin{split}\boldsymbol{\Psi}_{0} & =\left(\begin{array}{c}
\boldsymbol{\psi}_{b}^{\left(+\right)}\\
\boldsymbol{\psi}_{b}^{\left(-\right)}
\end{array}\right),\quad\boldsymbol{\Psi}_{2N}=\left(\begin{array}{c}
\boldsymbol{\psi}_{t}^{\left(+\right)}\\
\boldsymbol{\psi}_{t}^{\left(-\right)}
\end{array}\right),\\
\boldsymbol{\Psi}_{2n-1} & =\left(\begin{array}{c}
\boldsymbol{H}_{uv}\left(n-1\right)\\
\boldsymbol{E}_{vu}\left(n-1\right)
\end{array}\right)\quad\mathrm{for}\:1\leq n\leq N\,,\\
\boldsymbol{\Psi}_{2n} & =\left(\begin{array}{c}
\boldsymbol{E}_{vu}\left(n-1\right)\\
\boldsymbol{H}_{uv}\left(n\right)
\end{array}\right)\quad\mathrm{for}\:1\leq n\leq N-1.
\end{split}
\]
Equations (\ref{eq: mod Maxwell HE and EH}) and (\ref{eq: waveguide representation}) yield the transfer matrices:
\begin{equation}
\begin{split}\boldsymbol{T}_{0} & =\left(\begin{array}{cc}
\boldsymbol{U}_{b} & \boldsymbol{U}_{b}\\
\frac{i\boldsymbol{M}_{bEH}\boldsymbol{U}_{b}}{h_{b}^{2}}\frac{1}{1-\boldsymbol{\theta}_{b}^{-1}} & \frac{i\boldsymbol{M}_{bEH}\boldsymbol{U}_{b}}{h_{b}^{2}}\frac{1}{1-\boldsymbol{\theta}_{b}}
\end{array}\right),\\
\boldsymbol{T}_{2n-1} & =\left(\begin{array}{cc}
\boldsymbol{0} & \boldsymbol{1}\\
\frac{h_{0}\left(n-1\right)}{h_{0}\left(n\right)} & \frac{i\boldsymbol{M}_{HE}\left(n-1\right)}{h_{0}\left(n\right)h_{1}\left(n-1\right)}
\end{array}\right),\:\\
\boldsymbol{T}_{2n} & =\left(\begin{array}{cc}
\boldsymbol{0} & \boldsymbol{1}\\
\frac{h_{1}\left(n-1\right)}{h_{1}\left(n\right)} & \frac{i\boldsymbol{M}_{EH}\left(n\right)}{h_{1}\left(n\right)h_{0}\left(n\right)}
\end{array}\right),\:\quad1\leq n\leq N-1.
\end{split}
\label{eq: transfer matrix}
\end{equation}
For $\boldsymbol{T}_{2N-1}$, we have
\begin{equation}
\begin{split}\boldsymbol{T}_{2N-1} & =\left(\begin{array}{cc}
\boldsymbol{0} & -ih_{t}^{2}\left(1-\boldsymbol{\theta}_{t}^{-1}\right)\boldsymbol{U}_{t}^{\mathrm{T}}\\
\boldsymbol{1} & ih_{t}^{2}\boldsymbol{U}_{t}\left(1-\boldsymbol{\theta}_{t}^{-1}\right)\boldsymbol{U}_{t}^{\mathrm{T}}
\end{array}\right).\end{split}
\label{eq: T 2N-1 half}
\end{equation}
By using Eqs. (\ref{eq: transfer matrix}) and (\ref{eq: T 2N-1 half}), we can obtain the $2LM\times2LM$ matrices $\hat{\boldsymbol{t}}$  and $\hat{\boldsymbol{r}}$ from the linear equation:
\begin{equation}
\begin{split}\left(\begin{array}{c}
\hat{\boldsymbol{t}}\\
\boldsymbol{0}
\end{array}\right) & =\boldsymbol{T}_{2N-1}\cdots\boldsymbol{T}_{0}\left(\begin{array}{c}
\boldsymbol{1}\\
\hat{\boldsymbol{r}}
\end{array}\right),\end{split}
\label{eq: t-r equation}
\end{equation}
which is the same as Eq. (2.17) in our previous study~\cite{Usuki}.

\subsection{Stable transfer matrix method}

To solve the above Eq. (\ref{eq: t-r equation}), we can establish a stable
iterative procedure~\cite{Usuki, Akis} by using the $4LM\times4LM$ column operator $\boldsymbol{P}_{j}$.
This procedure does not entail solving multi-slice eigenvalue problems for the region with scattering and absorption, which has advantages in both computational time and numerical precision over the other procedure~\cite{Ko, Usuki94} that can be applied to optical scattering~\cite{Tikhodeev,Li,Liscidini,Anttu}.
In the following discussion, suppose we have $2LM\times2LM$ blocks of matrices $\mathcal{M}$ and $\mathcal{N}$ notated by
\[
\mathcal{M}=\left(\begin{array}{cc}
\mathcal{M}_{00} & \mathcal{M}_{01}\\
\mathcal{M}_{10} & \mathcal{M}_{11}
\end{array}\right)\quad \mathrm{and} \quad
\mathcal{N}=\left(\begin{array}{cc}
\mathcal{N}_{00} & \mathcal{N}_{01}
\end{array}\right).
\]
 The iterative equations for the $4LM\times4LM$ matrix $\boldsymbol{C}_{k}$
and the $2LM\times4LM$ matrix $\boldsymbol{D}_{k}$ can be used to find $\hat{\boldsymbol{t}}$
and $\hat{\boldsymbol{r}}$: 
\begin{equation}
\begin{split}
\boldsymbol{C}_{k+1} & = \boldsymbol{T}_{k}\boldsymbol{C}_{k}\boldsymbol{P}_{k},\\
\boldsymbol{D}_{k+1} & = \boldsymbol{D}_{k}\boldsymbol{P}_{k}
\quad \mathrm{for} \quad
0 \leq k \leq 2N-1.
\end{split}
\label{eq: C and D iterations}
\end{equation}
The initial conditions are that
\begin{equation}
\boldsymbol{C}_{0} = \left(\begin{array}{cc}
\boldsymbol{1} & \boldsymbol{0}\\
\boldsymbol{0} & \boldsymbol{1}
\end{array}\right)
\quad \mathrm{and} \quad
\boldsymbol{D}_{0}=\left(\begin{array}{cc}
\boldsymbol{0} & \boldsymbol{1}\end{array}\right).
\label{eq: C and D initial condition}
\end{equation}
$\boldsymbol{C}_{k}$ always satisfies
\begin{equation*}
\boldsymbol{C}_{k} = \left(
\begin{array}{cc}
\boldsymbol{C}_{k,00} & \boldsymbol{C}_{k,01}\\
\boldsymbol{0} & \boldsymbol{1}
\end{array}
\right)
\end{equation*}
because of the column operator $\boldsymbol{P}_{k}$ in Eq. (\ref{eq: C and D iterations}).
$\boldsymbol{P}_{k}$ has the following matrix representation:
\begin{eqnarray}
\boldsymbol{P}_{k} & = & \left(
\begin{array}{c}
\boldsymbol{1}\\
-\frac{ \boldsymbol{1}}{\boldsymbol{T}_{k,10} \boldsymbol{C}_{k,01} + \boldsymbol{T}_{k,11} } \boldsymbol{T}_{k,10} \boldsymbol{C}_{k,00}
\end{array}
\right.\nonumber \\
 &  & \qquad \qquad \qquad \left.
\begin{array}{c}
\boldsymbol{0}\\
\frac{\boldsymbol{1}}{ \boldsymbol{T}_{k,10} \boldsymbol{C}_{k,01} + \boldsymbol{T}_{k,11} }
\end{array}
\right),\label{eq: Pj}
\end{eqnarray}
and here, the actual numerical procedure uses Gaussian elimination {\it without} partial pivoting for $\boldsymbol{P}_{k11}$.
From Eqs. (\ref{eq: C and D initial condition}) and (\ref{eq:  Pj}), we find that iterating Eq. (\ref{eq: C and D iterations}) gives us 
\[
\hat{\boldsymbol{t}}=\boldsymbol{C}_{2N,00}\quad\mathrm{and}\quad\hat{\boldsymbol{r}}=\boldsymbol{D}_{2N,00}\,.
\]
We can compute the electromagnetic field in the scattering region by making $2LM\times4LM$ matrices $\hat{\mathcal{E}}\left(n,\, k\right)$ for $\boldsymbol{E}_{vu}$ and $\hat{\mathcal{H}}\left(n,\, k\right)$ for $\boldsymbol{H}_{uv}$ for the $n$-th cell.
The initial conditions are 
\[
\hat{\mathcal{E}}\left(n,\,2n+1\right) = \hat{\mathcal{H}}\left(n,\,2n\right) =\left(
\begin{array}{cc}
\boldsymbol{0} & \boldsymbol{1}
\end{array}
\right)
\]
for $1\leq n\leq N-2$. The $\hat{\mathcal{E}}\left(n,\, k\right)$
and $\hat{\mathcal{H}}\left(n,\, k\right)$ are iterated using the column operator $\boldsymbol{P}_{k}$: 
\[
\begin{split}
\hat{\mathcal{E}}\left(n,\, k+1\right) & =\hat{\mathcal{E}}\left(n,\, k\right)\boldsymbol{P}_{k}\quad\mathrm{for}\:2n+1\leq k\leq2N-1,\\
\hat{\mathcal{H}}\left(n,\, k+1\right) & =\hat{\mathcal{H}}\left(n,\, k\right)\boldsymbol{P}_{k}\quad\mathrm{for}\:2n\leq k\leq2N-1.
\end{split}
\]
\\ \indent
Appendices \ref{sec: Appendix Reverse_iteration I} and \ref{sec: Appendix Reverse_iteration II} give two approaches to simulating the reverse scattering process from the top ideal waveguide to the bottom ideal waveguide.
We can use either of these approaches to obtain $\hat{\boldsymbol{t}}^{'}$, $\hat{\boldsymbol{r}}^{'}$, $\hat{\mathcal{E}}^{'}\left(n,\, k\right)$ and $\hat{\mathcal{H}}^{'}\left(n,\, k\right)$ for the reverse process.

\subsection{Scattering matrix}

Let us define a $J_{b}\times J_{b}$ matrix $\boldsymbol{r}$ and a $J_{t}\times J_{b}$ matrix $\boldsymbol{t}$ only for propagating wave modes.
The elements of $\boldsymbol{r}$ and $\boldsymbol{t}$
are normalized to $\hat{\boldsymbol{r}}$ and $\hat{\boldsymbol{t}}$ by the power flow of Eq. (\ref{eq: gamma}) as follows:
\[
r_{jj^{'}}=\frac{\gamma_{b}\left(j\right)}{\gamma_{b}\left(j^{'}\right)}\hat{r}_{jj^{'}}\,,\quad t_{jj^{'}}=\frac{\gamma_{t}\left(j\right)}{\gamma_{b}\left(j^{'}\right)}\hat{t}_{jj^{'}}\,.
\]
If we only obtain the matrices $\boldsymbol{r}$ and $\boldsymbol{t}$,
we can reduce the matrix sizes of $\boldsymbol{C}_{k}$
and $\boldsymbol{D}_{k}$: $2LM\times J_{b}$ blocks
$\boldsymbol{C}_{k,00}$ and $\boldsymbol{C}_{k,10}$,
$2LM\times2LM$ blocks $\boldsymbol{C}_{k,01}$ and
$\boldsymbol{C}_{k,11}$, a $J_{b}\times J_{b}$
block $\boldsymbol{D}_{k,10}$, and a $J_{b}\times2LM$
block $\boldsymbol{D}_{k,11}$.
By using the matrices $\hat{\boldsymbol{r}}^{'}$ and $\hat{\boldsymbol{t}}^{'}$,
we can define a $J_{t}\times J_{t}$ matrix $\boldsymbol{r}^{'}$
and a $J_{b}\times J_{t}$ matrix $\boldsymbol{t}^{'}$. In so doing,
we obtain a $\left(J_{b}+J_{t}\right)\times\left(J_{b}+J_{t}\right)$
S-matrix~\cite{Buttiker}: 
\begin{equation}
\boldsymbol{S}=\left(\begin{array}{cc}
\boldsymbol{r} & \boldsymbol{t}^{'}\\
\boldsymbol{t} & \boldsymbol{r}^{'}
\end{array}\right)\,.\label{eq: S-matrix}
\end{equation}
Let us define $2LM\times J_{b}$ matrices $\mathcal{E}\left(n\right)$,
$\mathcal{H}\left(n\right)$
\begin{equation*}
\begin{split}
\left(\mathcal{E}\left(n\right)\right)_{jj^{'}} & =\frac{\left(\hat{\mathcal{E}}_{00}\left(n,2N\right)\right)_{jj^{'}}}{\gamma_{b}\left(j^{'}\right)},\\
\left(\mathcal{H}\left(n\right)\right)_{jj^{'}} &
=\frac{\left(\hat{\mathcal{H}}_{00}\left(n,2N\right)\right)_{jj^{'}}}{\gamma_{b}\left(j^{'}\right)}.
\end{split}
\end{equation*}
 and $2LM\times J_{t}$ matrices $\mathcal{E}^{'}$,
$\mathcal{H}^{'}$ only for propagating wave modes. 
We also define a $4LM\times\left(J_{b}+J_{t}\right)$
matrix $\boldsymbol{\xi}_{n}$ and a $4LM\times4LM$ matrix $\boldsymbol{\alpha}_{n}$
\begin{equation}
\begin{split}
\boldsymbol{\xi}_{n} & =  \left(\begin{array}{cc}
\mathcal{E}\left(n\right) & \mathcal{E}^{'}\left(n\right)\\
\mathcal{H}\left(n\right) & \mathcal{H}^{'}\left(n\right)
\end{array}\right),\\
\boldsymbol{\alpha}_{n} & =  \frac{c^{3}}{2 \omega^{2}} \left(\begin{array}{cc}
\mathrm{Im}\boldsymbol{M}_{HE}\left(n\right) & \boldsymbol{0}\\
\boldsymbol{0} & \mathrm{Im}\boldsymbol{M}_{EH}\left(n\right)
\end{array}\right).
\end{split}
\label{eq: xi alpha}
\end{equation}
From Eqs. (\ref{eq: power flow and dissipation}) and (\ref{eq: xi alpha}), the S-matrix including the case of absorption media satisfies
\begin{equation}
\boldsymbol{S}^{\dagger}\boldsymbol{S}-\boldsymbol{1}+\sum_{n=1}^{N-2}\boldsymbol{\xi}_{n}^{\dagger}\boldsymbol{\alpha}_{n}\boldsymbol{\xi}_{n}=\boldsymbol{0}.\label{eq: unitarity}
\end{equation}
This equation shows that the S-matrix is unitary when $\mathrm{Im}\boldsymbol{M}_{HE}\left(n\right)=\mathrm{Im}\boldsymbol{M}_{EH}\left(n\right)=0$ for $0\leq n\leq N-1$.

\section{Numerical results\label{sec:Numerical results}}

\begin{figure}[ht]
	\begin{centering}
		\includegraphics[width=0.9\columnwidth]{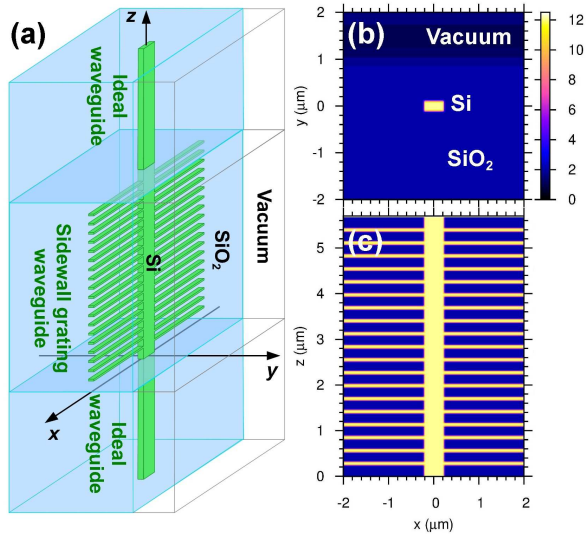}
	\end{centering}
	\caption{(a) SGW. (b) Permittivity distribution on $xy$-plane at $z=0$. (c) Distribution on $zx$-plane at $y=0$.}
	\label{fig:sidewall grating}
\end{figure}

\begin{figure}[ht]
\begin{centering}
\includegraphics[width=0.95\columnwidth]{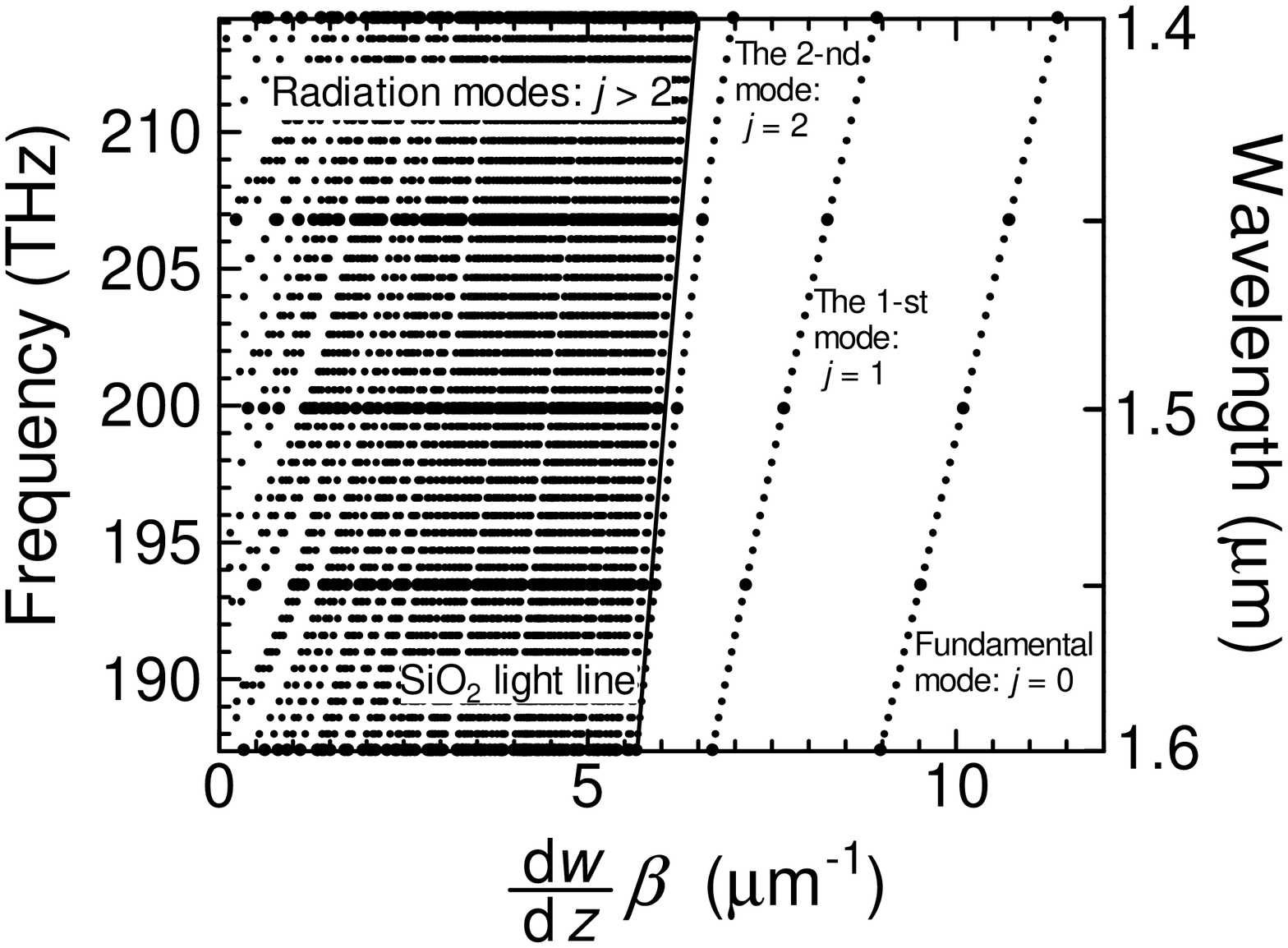}
\end{centering}
    \caption{Dispersion diagram of Si waveguide. There are three waveguide modes and many radiation modes in the $\rm{SiO_{2}}$ cladding layer and vacuum layer.}
    \label{fig:Dispersion}
\end{figure}

\begin{figure}[ht]
	\begin{centering}
		\includegraphics[width=0.95\columnwidth]{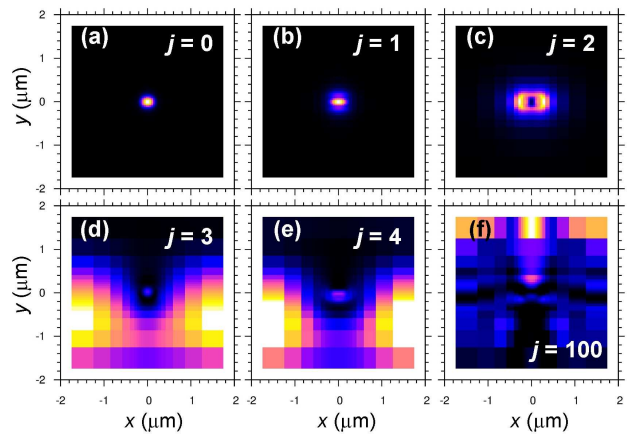}
	\end{centering}
	\caption{Optical modes for Si waveguide at 1.55 $\mu{\rm m}$.
		(a) $j=0$: $E^{x}_{11}$.
		(b) $j=1$: $E^{y}_{11}$.
		(c) $j=2$: $E^{x}_{21}$.
		(d) $j=3$: radiation mode in the $\rm{SiO_{2}}$ cladding layer.
		(e) $j=4$: radiation mode in the $\rm{SiO_{2}}$ cladding layer.
		(f) $j=100$: radiation mode in the vacuum layer.
	}
	\label{fig:optical mode}
\end{figure}

\begin{figure}[ht]
\begin{centering}
\includegraphics[width=0.9\columnwidth]{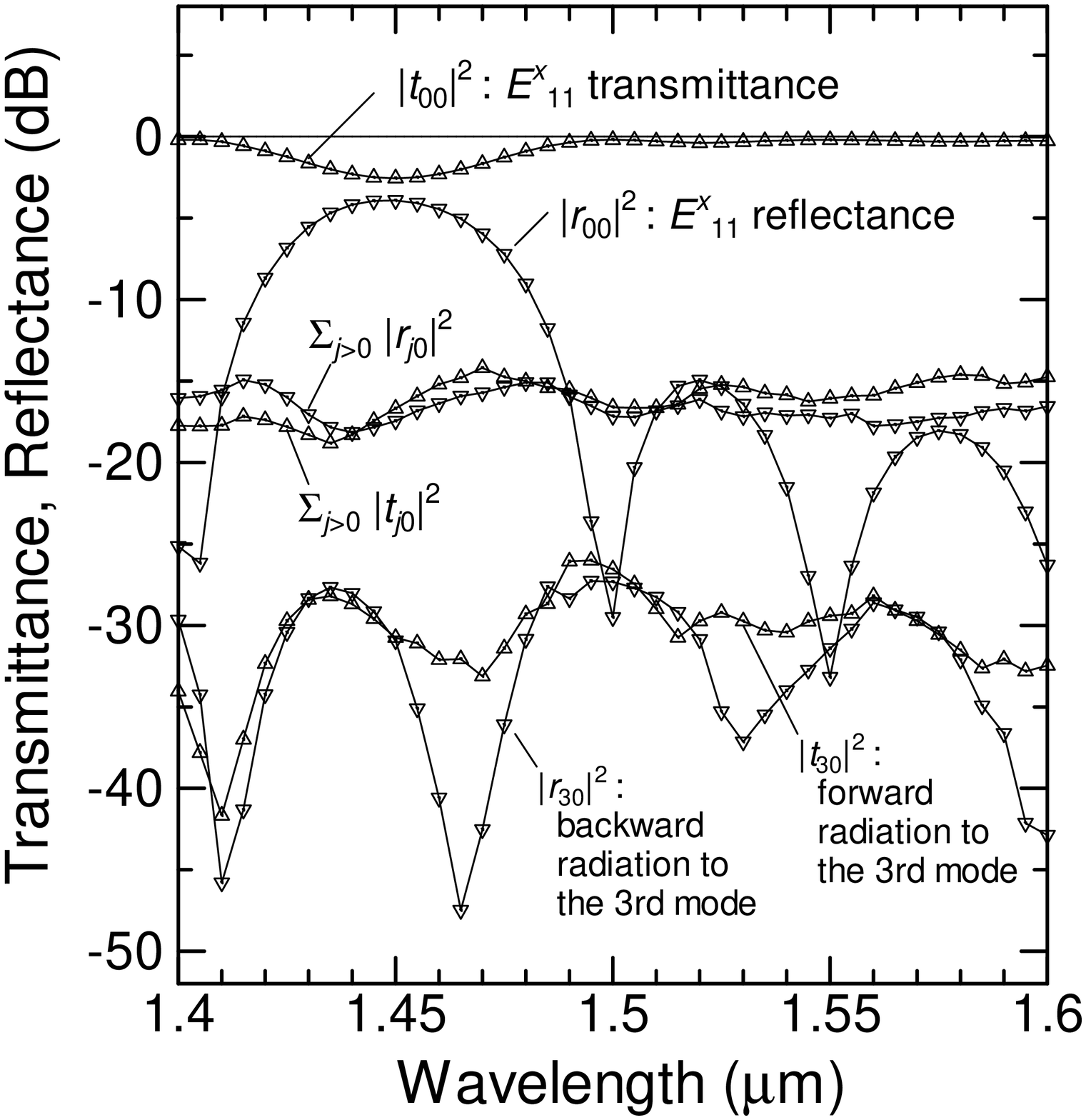}
\end{centering}
    \caption{Optical scattering from the fundamental mode. The SGW has a stop band at $1.45\,\mu{\rm m}$. This is a typical filtering property for $ \left | r_{00}\right |^{2}$, and it is designed for a modulator with low optical loss in the range of $1531 - 1591\,{\rm nm}$~\cite{CWDM}. The radiation and reflection loss is about $-15\,{\rm dB}$ respectively.}
    \label{fig:Radiations}
\end{figure}

\begin{figure}[ht]
\begin{centering}
\includegraphics[width=0.9\columnwidth]{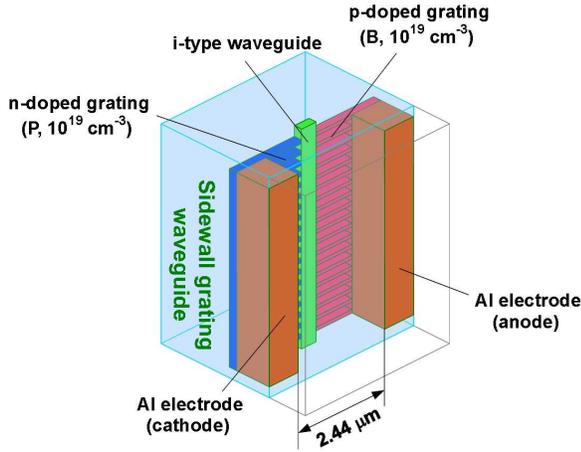}
\end{centering}
    \caption{P-i-n diode in SGW. The left (right) side of the side wall grating is doped with a donor (acceptor) density $10^{19}\,\rm{cm}^{-3}$. The electrodes' gap is $2.44\,\mu{\rm m}$. The permittivity of doped silicon is taken from~\cite{Soref, Reed}, and the permittivity of aluminum is taken from~\cite{Shiles, Ordal}. }
    \label{fig:Metal}
\end{figure}

\begin{figure}[ht]
\begin{centering}
\includegraphics[width=0.9\columnwidth]{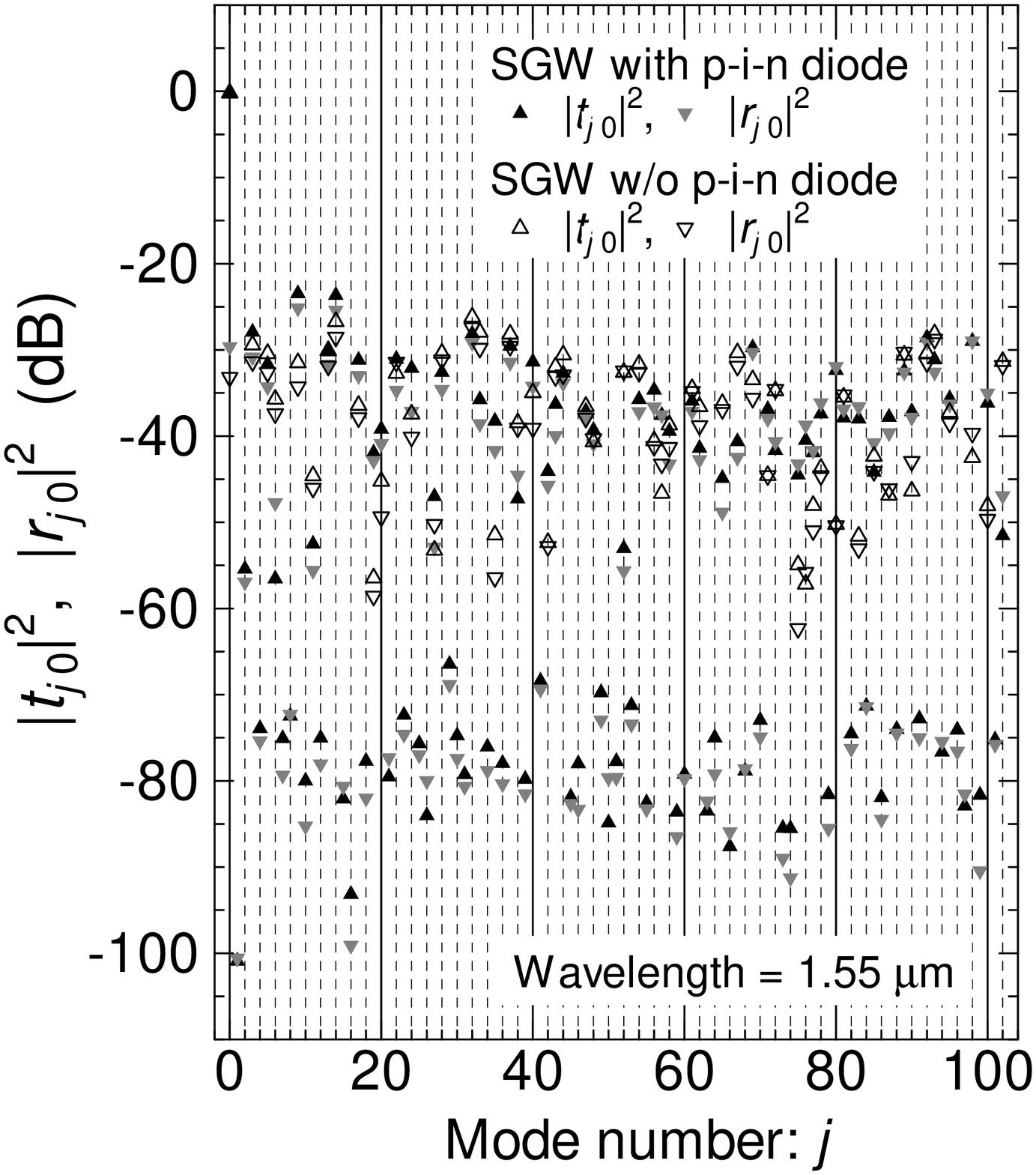}
\end{centering}
    \caption{Distribution of scattering coefficients on the SGW at $1.55{\mu}{\rm m}$. The distribution at around $-30\,{\rm dB}$ of the SGW with the p-i-n diode is not so different from that of the SGW without the diode.}
    \label{fig:absorbed_case}
\end{figure}

The method described in the previous section is suitable for analyzing very small scattering coefficients.
Here, we will discuss the optical properties of a sidewall grating waveguide (SGW) that is part of a phase shifter in a silicon optical modulator~\cite{Akiyama}.
The grating structure is used to inject free carriers into the waveguide core, but for it to work properly, the reflections of the fundamental mode and radiation loss from the structure have to be suppressed.
\\ \indent
Figure \ref{fig:sidewall grating} shows the SGW and its permittivity~\cite{Tong} distribution.
The SGW is a silicon waveguide  and has a $\rm{SiO_{2}}$ cladding layer on which a vacuum region is set.
For the numerical analysis, we set the waveguide core to $440\, \rm{nm} \times 220\, \rm{nm}$ and the grating pitch (width) to 284 (74.5) nm.
Before conducting the simulation, we have to obtain optical modes for the ideal waveguides at both ends of the SGW.
The ideal waveguides have three waveguide modes and many radiation modes, as shown in Fig. \ref{fig:Dispersion}.
The propagating mode numbers $J_{b}$ and $J_{t}$ are that $J_{b} = J_{t} = 125-96$ at $1.4-1.6\, \mu{\rm m}$,
because we set four $\mu{\rm m}$ periodic boundary conditions along the $x$ and $y$ axes. The grid parameters of $x$ and $y$ are the same as in Fig. \ref{fig: Grid}, whereas the grid parameters of $z$ are $dz/dw = 28.4\,\mathrm{nm}$ and $N = 200$.
Figure \ref{fig:optical mode} shows cross sections of the waveguide modes and radiation modes at a wavelength of 1.55 $\mu{\rm m}$.
The three waveguide modes in Fig. \ref{fig:optical mode}(a)-(c) are labeled $E^{x}_{11}$, $E^{y}_{11}$ and $E^{x}_{21}$~\cite{Okamoto}.
The waveguide modes are clearly localized around the silicon waveguide core.
From the $x$-axis symmetry of the SGW, the fundamental mode $E^{x}_{11}$ scattering through the SGW is intra-mode scattering (i.e. reflection) when the scattering only occurs between the waveguide modes.
However, the scattering process between the $E^{x}_{11}$ and radiation modes is complex.
Figure \ref{fig:optical mode}(d)-(f) shows three of the 100 radiation modes.
The optical power outside the core is dominant for each radiation mode, but  remains small inside the core.
Therefore, we have to consider not only the reflection within the $E^{x}_{11}$ but also scattering between $E^{x}_{11}$ and other modes including radiation modes.
\\ \indent
Figure \ref{fig:Radiations} shows the transmittance and reflectance of the SGW when $E^{x}_{11}$ is launched from the bottom.
There is a stop band at around $1.45\, \mu{\rm m}$, after which the fundamental-mode reflectance decreases with a periodic modification of increasing wavelength.
The reflectance of each radiation mode also has a strong wavelength dependence. In particular, it has a low value at the point of the third mode reflectance in Fig. \ref{fig:Radiations}.
The total optical loss, which is caused by $E^{x}_{11}$ reflection and radiation-mode scattering, is $4.39\,\%$ ($-13.6\,{\rm dB}$) at a wavelength of $1.55\,\mu{\rm m}$.
Note that the loss does not have so strong a wavelength dependence. 
The simulation results clearly show that the sidewall grating does not cause significant losses or reflections for the silicon optical interposer~\cite{Urino}.
\\ \indent
In an actual phase shifter~\cite{Akiyama}, the gratings on either side of the waveguide core are doped with donors or acceptors, and aluminum electrodes are attached to the gratings (see Fig. \ref{fig:Metal}).
Doped silicon~\cite{Soref, Reed} with metal electrodes~\cite{Shiles, Ordal} changes the optical index and absorption properties of the SGW.
We calculated the scattering process for absorption at a wavelength of $1.55\,\mu{\rm m}$.
Figure \ref{fig:absorbed_case} shows the distribution of $\left| t_{j0} \right|^{2}$ and $\left| t_{j0} \right|^{2}$ as $ 0 \leq j < 103 $.
The total optical loss is $5.59\,\%$ ($-12.5\,{\rm dB}$), and it consists of $E^{x}_{11}$ reflection (see the case of $j = 0$ in Fig. \ref{fig:absorbed_case}), $E^{x}_{21}$ scattering ($j = 2$), radiation-mode scattering ($ 3 \leq j < 103 $), and absorption loss ($0.52\,\%$ ($-22.8\,{\rm dB}$)).
Note that the distribution around $-80\,{\rm dB}$ and the $E^{x}_{21}$ scattering are caused by an $x$-axis asymmetry due to the different optical indexes of the p-doped and n-doped regions~\cite{Soref, Reed}.

\section{Conclusions\label{sec:Conclusion}}
\begin{figure}[ht]
	\begin{centering}
		\includegraphics[width=0.9\columnwidth]{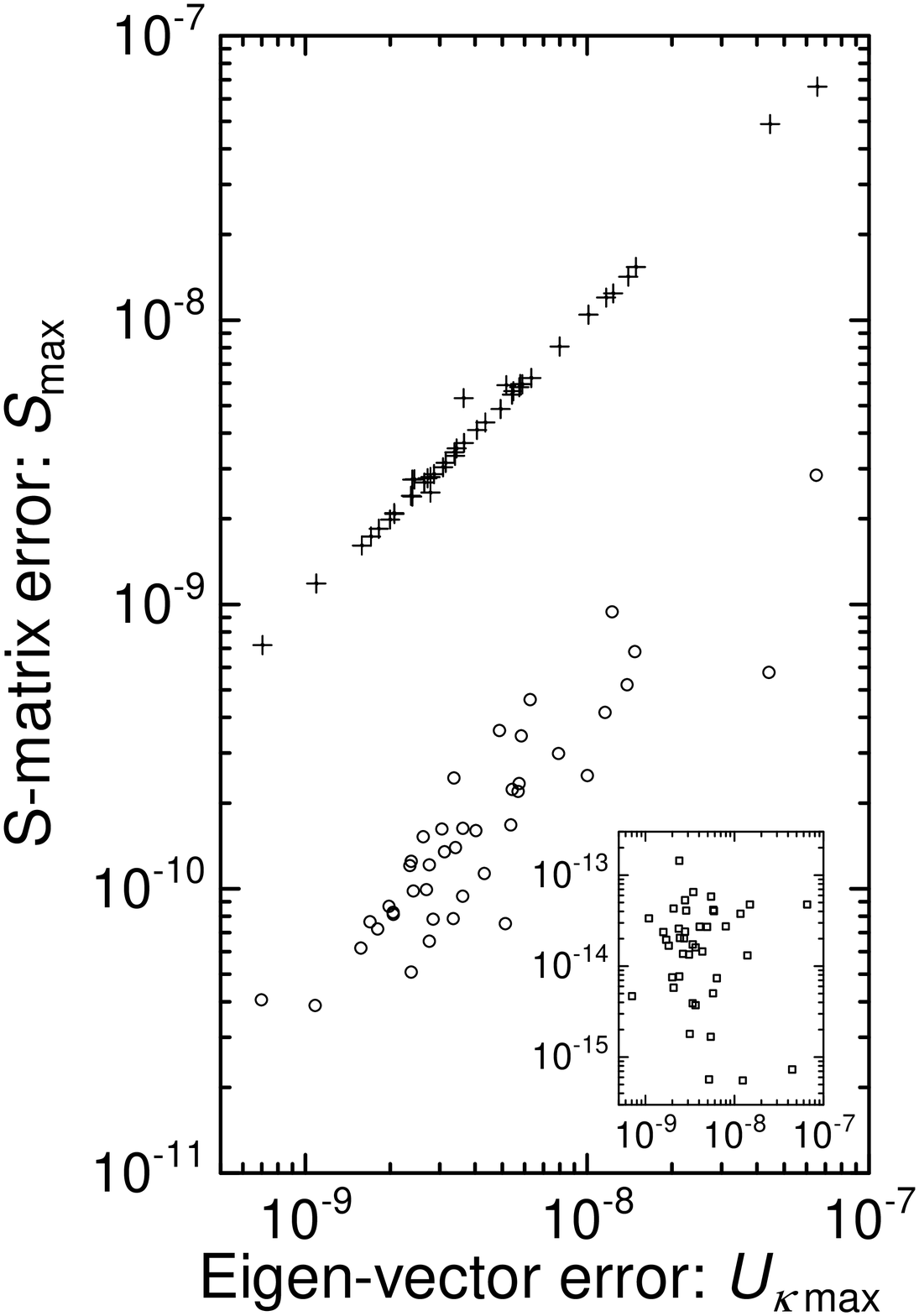}
	\end{centering}
	\caption{$U_{\kappa\,\max}$ dependence of $S_{\max}$. Circles (crosses) indicate Type I (II) results. $U_{b\,\max} = U_{t\,\max}$ in this case. The inset plots the absolute value of the $00$-element on the left side of Eq. (\ref{eq: orthogonality}).}
	\label{fig:Numerical-error}
\end{figure}

This proposed method can calculate scattering coefficients for all incident modes from the bottom waveguide and the top waveguide.
To determine the precision numerically, we use the max norm $S_{\max}$ on the left side of Eq. (\ref{eq: unitarity}):
\begin{eqnarray}
	S_{\max} = \left\Vert \boldsymbol{S}^{\dagger}\boldsymbol{S} - \boldsymbol{1}
			 + \sum_{n=1}^{N-2}\boldsymbol{\xi}_{n}^{\dagger}\boldsymbol{\alpha}_{n}\boldsymbol{\xi}_{n}\right\Vert _{\max}
	\label{eq: S_max}
\end{eqnarray}
and the max norm $U_{\kappa\,\max}$ on the left side of Eq. (\ref{eq: orthogonality}) for the propagation modes:
\begin{eqnarray}
	U_{\kappa\,\max} & = & \left \Vert \left(\boldsymbol{u}_{\kappa}\left(0\right)
	\cdots \boldsymbol{u}_{\kappa}\left(J_{\kappa}-1\right) \right)^{\dagger} \boldsymbol{M}_{\kappa  EH} \right. \nonumber \\
	&  & \times \left.
	\left(\boldsymbol{u}_{\kappa}\left(0\right)
	\cdots \boldsymbol{u}_{\kappa}\left(J_{\kappa}-1\right)\right)  - \boldsymbol{1} \right\Vert _{\max}.
	\label{eq: U_max}
\end{eqnarray}
We can estimate the numerical error of the obtained scattering coefficients by performing a double precision calculation of Eqs. (\ref{eq: S_max}) and (\ref{eq: U_max}).
Figure \ref{fig:Numerical-error} shows that numerical error of the eigenvalue calculation for ideal waveguide modes causes $S_{\max}$ error.
Using the Type I reverse propagation described in Appendix \ref{sec: Appendix Reverse_iteration I} results in an $S_{\max}$ value of less than $10^{-8}$.
On the other hand, the numerical error of the power-flow conservation for $E^{x}_{11}$ scattering does not depend on $U_{\kappa\,\max}$, and it is less than $2 \times 10^{-13}$, as shown in the inset of Fig. \ref{fig:Numerical-error}. 
Therefore, our method satisfies the condition placed on the S-matrix (Eq. (\ref{eq: S_max})) and gives us detailed optical properties with high enough precision for designing silicon photonics devices~\cite{Urino,Akiyama}.
For typical simulations, we recommend the Type II in Appendix \ref{sec: Appendix Reverse_iteration II}, because it takes only about half of the computational effort of Type I.
\\ \indent
We will use this method to analyze low and complex optical scattering cases. 
Furthermore, the discretization of the permittivity distribution on the Yee lattice is compatible with FDTD.
By combining this method and FDTD, numerical analyses with an optical propagation model can be made general and detailed.

\begin{acknowledgments}
The author would like to thank Makoto Okano for telling me about the FDTD method, Suguru Akiyama and Takeshi Baba for their fruitful discussions on Sec. \ref{sec:Numerical results} of this paper, and Motomu Takatsu for his helpful comments on Appendix \ref{sec: Appendix Orthogonality_of_eigenmode}.
This work was supported by the ``Funding Program for World-Leading
Innovative R\&D on Science and Technology (FIRST Program).''
\end{acknowledgments}

\appendix
\section{Non-uniform mesh for equations (\ref{eq: fj gj hj})\label{sec:Appendix Non_uniform_mesh}}
Here, we show the details of the coordinate transformation. The periodic
boundary conditions should be applied to $dx/du$ and $dy/dv$:
\[
	\frac{dx}{du} \left(u\right) = \frac{dx}{du} \left(u+L\right),\;
	\frac{dy}{dv} \left(v\right) = \frac{dy}{dv} \left(v+M\right).
\]
We introduce the periodic function $F\left(\xi,\, K\right)$ into $dx/du$ and $dy/dv$.
\begin{equation}
	\begin{split}
		F\left(\xi,\, K\right) & = 2^{2K-1} \cos^{2K} \left(\pi\xi\right)	\prod_{J=1}^{K-1} \frac{J}{K+J}\,.
	\end{split}
	\label{eq: function F}
\end{equation}
The function (\ref{eq: function F}) has several properties:
\begin{eqnarray*}
	F\left(\xi+1,\, K\right) & = & F\left(-\xi,\, K\right)=F\left(\xi,\, K\right),\\
	\max F\left(\xi,\, K\right) & = & F\left(0,\, K\right)=2^{2K-1}\prod_{J=1}^{K-1}\frac{J}{K+J},\\
	\min F\left(\xi,\, K\right) & = & F\left(\frac{1}{2},\, K\right)=0\,,\\
	\int_{0}^{1}F\left(\xi,\, K\right)d\xi & = & 1,\:\\
	\lim_{K\rightarrow\infty}F\left(\xi,\, K\right) & = & \sum_{I=-\infty}^{\infty}\delta\left(\xi-I\right),
\end{eqnarray*}
where $\delta\left(\xi\right)$ is the Dirac delta function.
From Eq. (\ref{eq: function F}), we obtain an analytical formula for the
integral of $F\left(\xi,\, K\right)$. 
\begin{equation}
	\begin{split}
		  & \int_{0}^{\xi}F\left(\xi^{'},\, K\right)d\xi^{'}\\
		= & \xi+\sum_{J=1}^{K}2\left(\prod_{J^{'}=1}^{J}\frac{K-J+J^{'}}{K+J^{'}}\right)\frac{\sin\left(2\pi J\xi\right)}{2\pi J}\,.
	\end{split}
	\label{eq: integral of F}
\end{equation}
The integral of $F$ has a staircase shape. 
For example, we can set $x\left(u\right)$ and $y\left(v\right)$ by using
\begin{equation}
	\begin{split}
		\frac{x\left(u\right)-x\left(0\right)-u\,\min\left(dx/du\right)}{x\left(L\right)-x\left(0\right)-L\,\min\left(dx/du\right)}
		& = \int_{0}^{u/L}F\left(\xi,\, K_{u}\right)d\xi,\\
		\frac{y\left(v\right)-y\left(0\right)-v\,\min\left(dy/dv\right)}{y\left(M\right)-y\left(0\right)-M\,\min\left(dy/dv\right)}
		& = \int_{0}^{v/M}F\left(\xi,\, K_{v}\right)d\xi,
	\end{split}
	\label{eq: xy example}
\end{equation}
given ten parameters $L$, $M$, $x\left(0\right)$, $y\left(0\right)$, $x\left(L\right)$, $y\left(M\right)$ $\min\left(dx/du\right)$,
$\min\left(dy/dv\right)$, $K_{u}$ and $K_{v}$.

\section{Orthogonality of eigenvalue equation (\ref{eq: eigenmode})
\label{sec: Appendix Orthogonality_of_eigenmode}}
We should note that $\boldsymbol{u}_{\kappa}^{\dagger}\left(j\right)\boldsymbol{M}_{\kappa EH}\,\boldsymbol{u}_{\kappa}\left(j\right)=0$,
where ``$^{\dagger}$'' denotes the Hermitian conjugate, when $\mathrm{Im}\Lambda_{\kappa}^{2}\left(j\right)\neq0$,
and that no linear combination $a\boldsymbol{M}_{\kappa EH}+b\boldsymbol{M}_{\kappa HE}^{-1}$
(real $a,\, b$) is positive definite.
However, the eigenvector $\boldsymbol{u}_{\kappa}$ still satisfies 
\[
\boldsymbol{u}_{\kappa}^{\mathrm{T}}\left(j\right)\boldsymbol{M}_{\kappa EH}\,\boldsymbol{u}_{\kappa}\left(j\right) \neq 0.
\]
As an example, let us consider a simple equation, $\left(\begin{array}{cc}
0 & 1\\
1 & 0
\end{array}\right)\boldsymbol{u}=\lambda\left(\begin{array}{cc}
1 & 0\\
0 & -1
\end{array}\right)\boldsymbol{u}$, where the matrix $\left(\begin{array}{cc}
b & a\\
a & -b
\end{array}\right)$ is not positive definite.
This equation has complex eigenvalues and eigenvectors: $\lambda_{\pm}=\pm i$, $\boldsymbol{u}_{\pm}=\left(\begin{array}{c}
1\\
\pm i
\end{array}\right)$, and
\[
\boldsymbol{u}_{\pm}^{\mathrm{T}}\left(\begin{array}{cc}
1 & 0\\
0 & -1
\end{array}\right)\boldsymbol{u}_{\pm} \neq 0.
\]
\indent
Therefore, we can normalize all eigenvectors as
\[
\boldsymbol{u}_{\kappa}^{\mathrm{T}}\left(j\right)\boldsymbol{M}_{\kappa EH}\boldsymbol{u}_{\kappa}\left(j'\right)=\delta_{jj^{'}}
\]
when eigenvalues are non-zero and non-degenerate, and number of eigenvalues is equal to the order of $\boldsymbol{M}_{\kappa EH}$. 
The eigenvalue equations for the numerical results of Sec. \ref{sec:Numerical results} satisfy this condition.
\\ \indent
We recommend checking which eigenvalue equations satisfy the condition or not, because there exists a counter example: $\left(\begin{array}{cc}
2 & 1\\
1 & 0
\end{array}\right)\boldsymbol{u}=\lambda\left(\begin{array}{cc}
1 & 0\\
0 & -1
\end{array}\right)\boldsymbol{u}$.
This equation has only one linearly independent eigenvector $\boldsymbol{u}=\left(\begin{array}{c} 1\\ -1 \end{array}\right)$ and
\[
\boldsymbol{u}^{\mathrm{T}}\left(\begin{array}{cc}
2 & 1\\
1 & 0
\end{array}\right)\boldsymbol{u}=0.
\]
Accordingly, we should carefully normalize eigenvectors when the equation  does not satisfy the condition.

\section{Reverse propagation type I\label{sec: Appendix Reverse_iteration I}}
For reverse propagation in the same framework as that of forward propagation,
we define a $4LM\times1$ column vector $\boldsymbol{\Psi}_{k}^{'}$ as
\[
\begin{split}
	\boldsymbol{\Psi}_{k}^{'} & = \left(
	\begin{array}{cc}
		\boldsymbol{0} & \boldsymbol{1}\\
		\boldsymbol{1} & \boldsymbol{0}
	\end{array}
\right)\boldsymbol{\Psi}_{k} \quad \mathrm{for} \; 2N \geq k \geq 0\,.\end{split}
\]
The $4LM\times4LM$ transfer matrix $\boldsymbol{T}_{k}^{'}$, which
satisfies $ \boldsymbol{\Psi}_{k}^{'} = \boldsymbol{T}_{k}^{'} \boldsymbol{\Psi}_{k+1}^{'} $,
is defined as 
\[
	\boldsymbol{T}_{2N-1}^{'} = \left(
	\begin{array}{cc}
		\frac{i\boldsymbol{M}_{tEH}\boldsymbol{U}_{t}}{h_{t}^{2}}\frac{1}{1-\boldsymbol{\theta}_{t}} & \frac{i\boldsymbol{M}_{tEH}\boldsymbol{U}_{t}}{h_{t}^{2}} \frac{1}{1-\boldsymbol{\theta}_{t}^{-1}}\\
		\boldsymbol{U}_{t} & \boldsymbol{U}_{t}
	\end{array}
\right),
\]
for $N>n\geq1$,
\begin{eqnarray*}
\boldsymbol{T}_{2n}^{'} & = & \left(\begin{array}{cc}
\boldsymbol{0} & \boldsymbol{1}\\
\frac{h_{1}\left(n\right)}{h_{1}\left(n-1\right)} & \frac{-i\boldsymbol{M}_{EH}\left(n\right)}{h_{1}\left(n-1\right)h_{0}\left(n\right)}
\end{array}\right),\\
\boldsymbol{T}_{2n-1}^{'} & = & \left(\begin{array}{cc}
\boldsymbol{0} & \boldsymbol{1}\\
\frac{h_{0}\left(n\right)}{h_{0}\left(n-1\right)} & \frac{-i\boldsymbol{M}_{HE}\left(n-1\right)}{h_{0}\left(n-1\right)h_{1}\left(n-1\right)}
\end{array}\right),
\end{eqnarray*}
and%
\[
\begin{split}\boldsymbol{T}_{0}^{'} & =\left(\begin{array}{cc}
-ih_{b}^{2}\left(1-\boldsymbol{\theta}_{b}\right)\boldsymbol{U}_{b}^{\mathrm{T}} & \boldsymbol{0}\\
ih_{b}^{2}\boldsymbol{U}_{b}\left(1-\boldsymbol{\theta}_{b}\right)\boldsymbol{U}_{b}^{\mathrm{T}} & \boldsymbol{1}
\end{array}\right).\end{split}
\]
The reverse equation corresponding to the forward Eq. (\ref{eq: t-r equation}) is
\[
\left(\begin{array}{c}
\hat{\boldsymbol{t}}^{'}\\
\boldsymbol{0}
\end{array}\right)=\boldsymbol{T}_{0}^{'}\cdots\boldsymbol{T}_{2N-1}^{'}\left(\begin{array}{c}
\boldsymbol{1}\\
\hat{\boldsymbol{r}}^{'}
\end{array}\right).
\]
Reverse iteration is defined as 
\[
\begin{split}\boldsymbol{C}_{k}^{'} & =\boldsymbol{T}_{k}^{'}\boldsymbol{C}_{k+1}^{'}\boldsymbol{P}_{k}^{'},\\
\boldsymbol{D}_{k}^{'} & =\boldsymbol{D}_{k+1}^{'}\boldsymbol{P}_{k}^{'} \quad \mathrm{for} \quad 0 \leq k \leq 2N-1,
\end{split}
\]
with initial conditions, 
\[
	\boldsymbol{C}_{2N}^{'}=\left(
	\begin{array}{cc}
		\boldsymbol{1} & \boldsymbol{0}\\
		\boldsymbol{0} & \boldsymbol{1}
	\end{array}
	\right) \quad \mathrm{and} \quad \boldsymbol{D}_{2N}^{'} = \left(
	\begin{array}{cc}
		\boldsymbol{0} & \boldsymbol{1}
	\end{array}
	\right).
\]
The column operator $\boldsymbol{P}_{k}^{'}$ can be defined in the same
manner as Eq. (\ref{eq: Pj}); that is,
\begin{eqnarray*}
	\boldsymbol{P}_{k}^{'} & = & \left(
	\begin{array}{c}
		\boldsymbol{1}\\
		- \frac{\boldsymbol{1}}{\boldsymbol{T}_{k,10}^{'} \boldsymbol{C}_{k+1,01}^{'} + \boldsymbol{T}_{k,11}^{'}}  \boldsymbol{T}_{k,10}^{'}\boldsymbol{C}_{k+1,00}^{'}
	\end{array}
	\right.\\
	 &  & \qquad\qquad\qquad\left.
	\begin{array}{c}
		\boldsymbol{0}\\
		\frac{\boldsymbol{1}}{\boldsymbol{T}_{k,10}^{'}\boldsymbol{C}_{k+1,01}^{'} + \boldsymbol{T}_{k,11}^{'}}
	\end{array}
	\right),
\end{eqnarray*}
without $k = 2N-1, \, 1$. $\boldsymbol{P}_{2N-1}^{'}$ and $\boldsymbol{P}_{1}^{'}$ are
\[
\boldsymbol{P}_{2N-1}^{'}=\boldsymbol{P}_{1}^{'}=\left(\begin{array}{cc}
\boldsymbol{1} & \boldsymbol{0}\\
\boldsymbol{0} & \boldsymbol{1}
\end{array}\right),
\]
because $\boldsymbol{T}_{2N-1}^{'}$ ($\boldsymbol{T}_{0}^{'}$) is different from $\boldsymbol{T}_{0}$ ($\boldsymbol{T}_{2N-1}$).
Here,
\begin{eqnarray*}
 &  & \boldsymbol{T}_{2N-2}^{'}\boldsymbol{C}_{2N-1}^{'}=\boldsymbol{T}_{2N-2}^{'}\boldsymbol{T}_{2N-1}^{'}\\
 & = & \left(\begin{array}{cc}
\boldsymbol{U}_{t} & \boldsymbol{U}_{t}\\
\frac{-i\boldsymbol{M}_{tEH}\boldsymbol{U}_{t}}{h_{1}\left(N-2\right)h_{t}}\frac{1}{1-\boldsymbol{\theta}_{t}^{-1}} & \frac{-i\boldsymbol{M}_{tEH}\boldsymbol{U}_{t}}{h_{1}\left(N-2\right)h_{t}}\frac{1}{1-\boldsymbol{\theta}_{t}}
\end{array}\right).
\end{eqnarray*}
Thus,
\[
\boldsymbol{P}_{2N-2}^{'}=\left(\begin{array}{cc}
\boldsymbol{1} & \boldsymbol{0}\\
\boldsymbol{\theta}_{t} & i \left(1-\boldsymbol{\theta}_{t}\right) h_{1}{\left(N-2\right)} h_{t}\boldsymbol{U}_{t}^{\mathrm{T}}
\end{array}\right)
\]
and
\begin{eqnarray*}
 &  & \boldsymbol{C}_{2N-2}^{'}=\boldsymbol{T}_{2N-2}^{'}\boldsymbol{C}_{2N-1}^{'}\boldsymbol{P}_{2N-2}^{'}\\
 & = & \left(\begin{array}{cc}
\boldsymbol{U}_{t}\left(1+\boldsymbol{\theta}_{t}\right)\, & \, i h_{1} {\left(N-2\right)} h_{t} \boldsymbol{U}_{t}\left(1-\boldsymbol{\theta}_{t}\right)\boldsymbol{U}_{t}^{\mathrm{T}}\\
\boldsymbol{0} & \boldsymbol{1}
\end{array}\right).
\end{eqnarray*}
Furthermore,
\begin{eqnarray*}
\boldsymbol{C}_{1}^{'} & = & \boldsymbol{T}_{1}^{'}\left(\begin{array}{cc}
\boldsymbol{C}_{2,00}^{'} & \boldsymbol{C}_{2,01}^{'}\\
\boldsymbol{0} & \boldsymbol{1}
\end{array}\right)\boldsymbol{P}_{1}^{'}\\
 & = & \left(\begin{array}{cc}
\boldsymbol{0} & \boldsymbol{1}\\
\boldsymbol{C}_{1,10}^{'} & \boldsymbol{C}_{1,11}^{'}
\end{array}\right).
\end{eqnarray*}
Thus,
\begin{eqnarray*}
\boldsymbol{T}_{0}^{'}\boldsymbol{C}_{1}^{'} & = & \left(\begin{array}{cc}
\boldsymbol{0} & -ih_{b}^{2}\left(1-\boldsymbol{\theta}_{b}\right)\boldsymbol{U}_{b}^{\mathrm{T}}\\
\boldsymbol{C}_{1,10}^{'} & \boldsymbol{C}_{1,11}^{'}+ih_{b}^{2}\boldsymbol{U}_{b}\left(1-\boldsymbol{\theta}_{b}\right)\boldsymbol{U}_{b}^{\mathrm{T}}
\end{array}\right).
\end{eqnarray*}
From Eqs. (\ref{eq: T 2N-1 half}), $\boldsymbol{T}_{0}^{'}\boldsymbol{C}_{1}^{'}$
is an expression similar to $\boldsymbol{T}_{2N-1} \boldsymbol{C}_{2N-1}$.
Reverse iteration yields
\[
\hat{\boldsymbol{t}}^{'}=\boldsymbol{C}_{0,00}^{'}\quad\mathrm{and}\quad\hat{\boldsymbol{r}}^{'}=\boldsymbol{D}_{0,10}^{'}\,.
\]
At the $n$-th cell, we can introduce $\hat{\mathcal{E}}^{'}\left(n,\,2N-k\right)$ and $\hat{\mathcal{H}}^{'}\left(n,\,2N-k\right)$.
For $1\leq n\leq N-2$, the initial conditions are
\[
	\hat{\mathcal{E}}^{'}\left(n,\,2N-2n-2\right)  = 
	\hat{\mathcal{H}}^{'}\left(n,\,2N-2n-1\right)  = \left(
	\begin{array}{cc}
		\boldsymbol{0} & \boldsymbol{1}
	\end{array}
	\right).
\]
The following iterations derive $\hat{\mathcal{E}}^{'}\left(n,\,2N\right)$
and $\hat{\mathcal{H}}^{'}\left(n,\,2N\right)$:
\[
\begin{split}\hat{\mathcal{E}}^{'}\left(n,\,2N-k\right) & =\hat{\mathcal{E}}^{'}\left(n,\,2N-k-1\right)\boldsymbol{P}_{k}^{'}\\
 & \qquad\qquad\mathrm{for} \, 2n+1\geq k \geq 0,\\
\hat{\mathcal{H}}^{'}\left(n,\,2N-k\right) & =\hat{\mathcal{H}}^{'}\left(n,\,2N-k-1\right)\boldsymbol{P}_{k}^{'}\\
 & \qquad\qquad\quad\mathrm{for} \, 2n \geq k \geq 0.
\end{split}
\]

\section{Reverse propagation type II\label{sec: Appendix Reverse_iteration II}}
We can derive another equation for $\hat{\boldsymbol{r}}^{'}$ and
$\hat{\boldsymbol{t}}^{'}$:
\begin{eqnarray*}
		\left(
		\begin{array}{c}
			\hat{\boldsymbol{r}}^{'}\\
			\boldsymbol{0}
		\end{array}
		\right) 
		& = &\boldsymbol{T}_{2N-1} \left[ \boldsymbol{T}_{2N-2}\cdots\boldsymbol{T}_{0} \left(
		\begin{array}{c}
			\boldsymbol{0}\\
			\hat{\boldsymbol{t}}^{'}
		\end{array}
		\right) \right.\\
		& &\qquad\qquad\qquad - \left. \left(
		\begin{array}{c}
			\boldsymbol{U}_{t}\\
			\frac{i\boldsymbol{M}_{tEH}\boldsymbol{U}_{t}}{h_{t}^{2}} \frac{1}{1-\boldsymbol{\theta}_{t}}
		\end{array}
		\right)\right].
\end{eqnarray*}
At the $2N-2\rightarrow2N-1$ step, we introduce $\boldsymbol{C}_{2N-1}^{'}$
and $\boldsymbol{D}_{2N-1}^{'}$:
\begin{eqnarray*}
	\boldsymbol{C}_{2N-1}^{'} & = & \left(
	\begin{array}{c}
		-\boldsymbol{U}_{t}+\boldsymbol{C}_{2N-1,01}\frac{i\boldsymbol{M}_{tEH}\boldsymbol{U}_{t}}{h_{t}^{2}}\frac{1}{1-\boldsymbol{\theta}_{t}}\\
		\boldsymbol{0}
	\end{array}
	\right.\\
	&  & \qquad\qquad\qquad\left.
	\begin{array}{c}
		\boldsymbol{C}_{2N-1,01}\\
		\boldsymbol{1}
	\end{array}
	\right)\,,\\
	\boldsymbol{D}_{2N-1}^{'} & = & \left(
	\begin{array}{cc}
		\boldsymbol{D}_{2N-1,01} \frac{i\boldsymbol{M}_{tEH}\boldsymbol{U}_{t}}{h_{t}^{2}} \frac{1}{1-\boldsymbol{\theta}_{t}}\, 
		& \boldsymbol{D}_{2N-1,01}
	\end{array}
	\right).
\end{eqnarray*}
Thus,
\begin{equation}
	\begin{split}
		\boldsymbol{C}_{2N}^{'} & =\boldsymbol{T}_{2N-1}\boldsymbol{C}_{2N-1}^{'}\boldsymbol{P}_{2N-1}^{'},\\
		\boldsymbol{D}_{2N}^{'} & =\boldsymbol{D}_{2N-1}^{'}\boldsymbol{P}_{2N-1}^{'}.
	\end{split}
	\label{eq: C' and D' iterations}
\end{equation}
Note that $\boldsymbol{C}_{k,01}^{'}=\boldsymbol{C}_{k,01}$, $\boldsymbol{D}_{k,01}^{'}=\boldsymbol{D}_{k,01}$
and $\boldsymbol{P}_{k,01}^{'}=\boldsymbol{P}_{k,01}$ for $2N-1 \leq k \leq2N$.
The iterative procedure of Eq. (\ref{eq: C' and D' iterations}) yields
\[
	\hat{\boldsymbol{r}}^{'}=\boldsymbol{C}_{2N,00}^{'}\quad\mathrm{and}\quad\hat{\boldsymbol{t}}^{'}=\boldsymbol{D}_{2N,00}^{'}\,.
\]
We can add $\hat{\mathcal{E}}^{'}\left(n,\,2N-1\right)$ and $\hat{\mathcal{H}}^{'}\left(n,\,2N-1\right)$
to the $2N-2\rightarrow2N-1$ step. 
\[
	\begin{split}
		\hat{\mathcal{E}}^{'}\left(n,\,2N-1\right) 
		& = \left(
		\begin{array}{cc}
			\boldsymbol{0} & \hat{\mathcal{E}}_{01} \left(n,\,2N-1\right)
		\end{array}\right),\\
		\hat{\mathcal{H}}^{'}\left(n,\,2N-1\right) 
		& =\left(
		\begin{array}{cc}
			\boldsymbol{0} & \hat{\mathcal{H}}_{01}\left(n,\,2N-1\right)
		\end{array}
		\right).
	\end{split}
\]
Finally, we obtain
\[
\begin{split}
	\hat{\mathcal{E}}^{'}\left(n,\,2N\right) & =\hat{\mathcal{E}}^{'}\left(n,\,2N-1\right)\boldsymbol{P}_{2N-1}^{'},\\
	\hat{\mathcal{H}}^{'}\left(n,\,2N\right) & =\hat{\mathcal{H}}^{'}\left(n,\,2N-1\right)\boldsymbol{P}_{2N-1}^{'}.
\end{split}
\]


\begin{thebibliography}{99}

\bibitem{Joannopoulos} J. D. Joannopoulos, S. G. Johnson, J. N. Winn, and R. D. Meade
``Photonic Crystals,'' 2nd-ed., Princeton University Press (2008).

\bibitem{Kimerling} L. C. Kimerling, D. Ahn, A. B. Apsel, M. Beals, D. Carothers, Y.-K. Chen, T. Conway, D. M. Gill, M. Grove, C.-Y. Hong, M. Lipson, J. Liu, J. Michel, D. Pan, S. S. Patel, A. T. Pomerene, M. Rasras, D. K. Sparacin, K.-Y. Tu, A. E. White, and C. W. Wong,
  Proc. SPIE \textbf{6125}, 612502-1 (2006).

\bibitem{Miller} D. A. B. Miller, Proc. IEEE \textbf{97}, 1166 (2009).

\bibitem{198008_taflove} A. Taflove, 
  IEEE Transactions on Electromagnetic Compatibility, \textbf{EMC-22}, 191 (1980).

\bibitem{Urino} Y. Urino, Y. Noguchi, M. Noguchi, M. Imai, M. Yamagishi, S. Saitou, N. Hirayama, M. Takahashi, H. Takahashi, E. Saito, M. Okano, T. Shimizu, N. Hatori, M. Ishizaka, T. Yamamoto, T. Baba, T. Akagawa, S. Akiyama, T. Usuki, D. Okamoto, M. Miura, J. Fujikata, D. Shimura, H. Okayama, H. Yaegashi, T. Tsuchizawa, K. Yamada, M. Mori, T. Horikawa, T. Nakamura, and Y. Arakawa,
 Opt. Express \textbf{20}, B256 (2012).

\bibitem{Yamamoto} T. Yamamoto, H. Kobayashi, M. Ekawa, S. Ogita, T. Fujii, T. Higashi and M. Kobayashi, Electronics Lett. \textbf{33}, 65 (1997).

\bibitem{Taflove_book} Chapter 4 in A. Taflove and S. C. Hagness, ``Computational Electrodynamics: The Finite-Difference Time-Domain Method,'' 2nd-ed., Artech House (2000).

\bibitem{IEEE754} IEEE, ``IEEE standard Floating-Point Arithmetic,'' IEEE Std 754-2008, pp. 1-58, Aug 2008.

\bibitem{Cho} K. Cho, ``Optical Response of Nano-structures: Microscopic Nonlocal Theory,'' Springer Verlag, Heidelberg (2003); Errata, Web site of Springer Verlag for this book.

\bibitem{Wheeler} J. A. Wheeler, Phys. Rev. \textbf{52}, 1107 (1937).

\bibitem{Buttiker} M. B{\"u}ttiker, Y. Imry, R. Landauer, and S. Pinhas,
Phys. Rev. B \textbf{31}, 6207 (1985).

\bibitem{Tikhodeev} S. G. Tikhodeev, A. L. Yablonskii, E. A. Muljarov, N. A. Gippius, and T. Ishihara,
Phys. Rev. B \textbf{66}, 045102 (2002).

\bibitem{Li} Z.-Y. Li and L.-L. Lin,
Phys. Rev. E \textbf{67}, 046607 (2003).

\bibitem{Liscidini} M. Liscidini, D. Gerace, L. C. Andreani, and J. E. Sipe,
Phys. Rev. B \textbf{77}, 035324 (2008).

\bibitem{Anttu} N. Anttu and H. Q. Xu,
Phys. Rev. B \textbf{83},165431 (2011).

\bibitem{Usuki} T. Usuki, M. Saito, M. Takatsu, R. A. Kiehl, and N. Yokoyama,
Phys. Rev. B \textbf{52}, 8244 (1995).

\bibitem{Akis} R. Akis and D. Ferry, J. Comput. Electron. \textbf{9}, 232 (2010).

\bibitem{Yee} K. S. Yee, 
IEEE Trans. Antennas Propag. \textbf{AP-14}, 302 (1966).

\bibitem{Okamoto} Equation (1.44) and Fig. 2.11 in K. Okamoto, ``Fundamentals of Optical Waveguides,'' 2nd-ed., Elsevier Inc. (2006).

\bibitem{Ko} D. Y. K. Ko and J. C. Inkson,
Phys. Rev. B \textbf{38}, 9945 (1988).

\bibitem{Usuki94} T. Usuki, M. Takatsu, R. A. Kiehl, and N. Yokoyama,
Phys. Rev. B \textbf{50}, 7615 (1994).

\bibitem{Akiyama} S. Akiyama, T. Baba, M. Imai, T. Akagawa,
	M. Takahashi, N. Hirayama, H. Takahashi, Y. Noguchi, H. Okayama, T. Horikawa, and T. Usuki
Opt. Express \textbf{20}, 2911 (2012).

\bibitem{Tong} Equations (6) and (7) in  L. Tong, J. Lou, and E. Mazur,
Opt. Express \textbf{12}, 1025 (2004).

\bibitem{CWDM}``ITU-T G.694.2,'' http://www.itu.int/rec/T-REC-G.694.2-200312-I/en

\bibitem{Soref} R. A. Soref, and B. R. Bennett,
IEEE J. Quantum Electron., \textbf{QE23}, 123 (1987).

\bibitem{Reed} G. T. Reed, G. Mashanovich, F. Y. Gardes and D. J. Thomson,
nature photonics \textbf{4}, 518 (2010).

\bibitem{Shiles} E. Shiles,  T. Sasaki, M. Inokuti and D. Y. Smith,
Phys. Rev. B \textbf{22}, 1612 (1980).

\bibitem{Ordal}Table 1 in  M.A. Ordal, L. L. Long, R. J. Bell, S. E. Bell, R. R. Bell, R. W. Alexander, Jr. and C. A. Ward,
Appl. Opt. \textbf{22}, 1099 (1983).

\end{thebibliography}
\end{document}